\documentclass[preprint]{aastex}
\usepackage{psfig}

\newcommand{\Bg}{Br$\gamma$}
\newcommand{\Ls}{L$_{\odot}$}
\newcommand{\Ms}{M$_{\odot}$}

\newcommand{\Myr}{M$_{\odot}$~yr$^{-1}$}
\newcommand{\kms}{km~s$^{-1}$}

\newcommand{\myvspace}{\rule[0pt]{0pt}{0.4cm}}

\slugcomment{To appear in The Astrophysical Journal}

\shorttitle{Multiwavelength Study of NGC 7714}
\shortauthors{Lan\c{c}on et al.}

\begin{document}
\title{Multiwavelength Study of the Starburst Galaxy NGC~7714.  \\
    II. The Balance between Young, Intermediate Age and Old Stars }

\author{Ariane Lan\c{c}on}
\affil{Obs. de Strasbourg (UMR 7550), 
	11 rue de l'Universit\'e, 67000 Strasbourg, France\\
       lancon@astro.u-strasbg.fr}

\author{Jeffrey D. Goldader}
\affil{Univ. of Pennsylvania, Dept. of Phys. \&
       Astron., 209 S. 33rd St., Philadelphia, PA~19104\\
       jdgoldad@dept.physics.upenn.edu}

\author{Claus Leitherer}
\affil{Space Telescope Science Institute\altaffilmark{1}, 
       3700 San Martin Drive,
       Baltimore, MD 21218\\
       leitherer@stsci.edu}

\and

\author{Rosa M. Gonz\'alez Delgado}
\affil{Instituto de Astrof\'{\i}sica de Andaluc\'{\i}a (CSIC), Apdo. 3004,
       18080 Granada, Spain\\
       rosa@iaa.es}

\altaffiltext{1}{Operated by AURA, Inc., under NASA contract NAS5-26555}


\begin{abstract}
We combine existing multiwavelength data (including an
HST/GHRS UV spectrum and a ground based optical spectrum) with 
unpublished HST/WFPC2 images, near-IR photometry and K band 
spectroscopy. We use these data to constrain the
young, the intermediate age and the old stellar populations in the
central regions of the starburst galaxy NGC\,7714.

In a previous paper (Gonz\'alez Delgado et al. 1999),
the stellar features in the HST/GHRS ultraviolet (UV) spectrum and the
optical emission lines were used to identify a
$\sim$\,5\,Myr old, very little reddened stellar population as
the main source of UV light in the central $\sim$330\,pc. The 
optical data indicated the existence of an older population. The
nature of the latter is investigated here. Stellar absorption
features in the optical and the near-IR are used to
partly break the strong degeneracy between the effects of 
ageing and those of the inhomogeneous dust distribution on the UV--optical--IR
colors. Consistency with far-IR, X-ray and radio data is also 
addressed. The successful models have essential features in common.
We find that the young burst responsible for the UV light represents
only a small part of an extended episode of enhanced star formation,
initiated a few 10$^8$\,yrs ago. The star formation rate is likely to have
varied on this timescale, averaging about 1\,\Ms\,yr$^{-1}$. The mass of young
and intermediate age stars thus formed equals at least 10\,\% of the
mass locked in pre-existing stars of the underlying spiral galaxy nucleus, 
and fractions around 25\,\% are favored. The spectrophotometric 
star formation timescale is long compared to the $\sim 110$\,Myr
elapsed since closest contact with the neighboring NGC\,7715, according
to the dynamical models of Smith \& Wallin (1992). The initial trigger of 
the starburst thus remains elusive.  

NGC\,7714 owes its brightness in the UV to a few low extinction lines of 
sight towards young stars. Our results based on the integrated 
spectrophotometry of the central $\sim\,330$\,pc are supported
by high resolution images of this area. The different extinction
values obtained when different spectral indicators are used 
result naturally from the coexistence of populations with various
ages and obscurations. The near-IR continuum image looks smoothest, as
a consequence of lower sensitivity to extinction and of a larger
contribution of old stars.
  
We compare the nuclear properties of NGC\,7714 with results from studies
in larger apertures. We emphasize that the global properties of 
starburst galaxies are the result of the averaging over many 
lines of sight with very diverse properties in terms
of obscuration and stellar ages. The overal picture is strongly 
reminiscent of the other nearby ``proto-typical" starburst, M\,82.
\end{abstract}

\keywords{dust, extinction---galaxies: individual 
         (NGC~7714)---galaxies: ISM---galaxies: starburst---galaxies: stellar
          content}


\section{Introduction}
\label{Intro.sec}

A burst of star formation in a galaxy affects the galaxy's energy output
across the entire electromagnetic spectrum. Supernovae emit X-rays; the
continua of hot massive stars are strong in the ultraviolet (UV); gaseous
recombination lines dominate the optical spectra; cool stars are
strong emitters in the near-infrared (near-IR); dust heated by the absorption
of energetic photons can produce strong far-IR emission; and synchrotron
radiation from electrons accelerated by supernova remnants is important
at radio wavelengths. Yet, the physical manifestations of a starburst 
depend on its age. Old starbursts ($\ga$ tens of Myr), where the
majority of the massive stars have evolved off the main sequence or already
died, will have relatively little UV emission, though red supergiants 
(RSG), red giants, or asymptotic giant branch (AGB) stars could cause them
to be quite bright in the near-IR. On the other hand, a very young starburst
($\la$ few Myr) will have strong UV emission, yet relatively weak IR
emission, since IR luminous stars have not yet formed. For a recent review
see, e.g., \citet{lei00}.

The effects of starbursts have been studied in great detail for large 
samples at individual wavelengths. Multiwavelength
studies of starbursts at galaxy-scale resolution were done by, e.g.,
\citet{cal97}, \citet{sch97}, and \citet{mas99}.
But few studies have attempted a panchromatic approach that
combines high spatial resolution with both photometric and 
spectroscopic information, focussing on one object, thereby 
studying star formation in an individual galaxy at the greatest possible
detail. 

One motivation for programs aiming at very high spatial resolution is 
disentangling the complex effects of reddening by dust (with various extinction 
geometries) from ``secular" age-induced reddening as stellar populations age. 
For all dusty starbursts, the reddening correction is an essential step in
the determination of the total amount of star formation. The conversion
of reddening measurements into attenuation factors is non trivial,
even for ``simple" stellar populations (Witt \& Gordon 2000).
For instance, it has often been suggested that emission lines 
are affected by more extinction than stellar continuum emission
(e.g. Calzetti et al. 1994). Only the detailed analysis of
spectrophotometric properties in the light of high resolution
imaging data will allow us to understand what is really happening
and to gain confidence in results for more distant objects, for
which only integrated spectrophotometry is available.

We have chosen to study the prototypical starburst galaxy NGC~7714
\citep{wee81}. This spiral galaxy at a distance of 37.3~Mpc (for 
$H_{\rm0} = 75$~\kms~Mpc$^{-1}$ and $cz = 2798$~\kms)
is interacting with its smaller neighbor NGC~7715 (see references in
Gonz\'alez Delgado et al. 1999,
hereafter Paper~I). The inclination of NGC~7714 is $\sim$45$^\circ$, allowing a 
fairly clear view of the inner few hundred pc ($1\arcsec$ is equivalent to
189~pc at 37.3~Mpc), where the strongest star formation is occurring.
Though luminous in the IR ($L_{\rm{IR}} = 3 \times 10^{10}$~\Ls; see Paper~I),
NGC~7714 is a strong UV source as well. Together, these facts are evidence
that the mean dust obscuration is not too severe. This gave us hope that
we could use data from the spacecraft UV to model directly the UV continuum
of the hot, young stars, providing powerful constraints on the starburst
age. This was the major topic of Paper~I.

The {\em global} properties of NGC~7714 were studied previously by 
\citet{cal97}, who analyzed large-aperture ($10\arcsec \times 20\arcsec$, 
1.9~kpc~$\times$~3.8~kpc) multiwavelength photometry and spectroscopy
similar to the data in our study. However, we concentrate on the inner
few hundred pc of the nucleus, where the most intense star formation is
occurring.

As part of Hubble Space Telescope (HST) Guest Observer program 6672, we
obtained a UV spectrum (1180~--~1680~\AA\ in the rest frame) of the
nucleus of NGC~7714 using the Goddard High Resolution Spectrograph
(GHRS). Our analysis of the UV spectrum and a comparison with available
X-ray, optical, and radio data was reported in Paper~I. We also included
in that paper an F606W image obtained with the Wide-Field Planetary
Camera~2 (WFPC2) on the HST, taken from the HST archive.
 
In this paper, we extend our analysis into the IR. This
wavelength range is particularly sensitive to older, evolved stars, and
to stars of all ages with heavy dust obscuration. We have obtained
near-IR JHK$n$ images, and a K-band spectrum, which we present and 
analyze here. We also present a new near-UV image of NGC~7714 taken
with HST. By combining the UV data from HST with optical and near-IR
spectroscopy from the ground, we have accumulated high-quality spectra,
spanning the range 1200~\AA\ to 2.3~$\mu$m, for very nearly the same
spatial regions of the galaxy. With photometric points from the X-ray
to the radio, our spatial coverage spans several decades in frequency.

The new observations are presented in Sect.\,\ref{obs.sec}. In
Sect.\,\ref{data.anal.sec}, the data are analysed with a focus
on the morphological and structural information they contain
for the central regions of NGC\,7714. The coexistence of regions
with very different properties, in particular in terms of extinction,
already becomes evident in that section. The integrated spectrophotometric
properties of the nucleus are analyzed in Sect.\,\ref{Spectrum.sec}.
We successively consider individual wavelength ranges, the full
broad band energy distribution, and finally the full spectrum,
and show that a variety of models remains consistent with even this amount
of combined data. The predictions common to all successful models 
are highlighted and analyzed in terms of the nuclear
morphology of the galaxy. Implications of our study for NGC\,7714 itself and 
for studies of other starbursts are discussed in Sect.\,\ref{Discussion.sec}.
The conclusions are given in Sect.\,\ref{Concl.sec}.

\section{Observations and data reduction}
\label{obs.sec}

\subsection{Spectroscopy at 2 $\mu$m}
\label{Kspec.obs.sec}

NGC~7714 was observed on the nights of 1996 July 26 and 28 UT, using
the CGS4 spectrometer on the United Kingdom Infrared Telescope.  
CGS4 was equipped with a 256$\times$256 InSb array.
We used the 150~mm focal length camera with the 150 line/mm grating in
first order.  The slit was one pixel (1.22$\arcsec$) wide.  
In this configuration, four
undersampled spectra were taken with the detector moved 0.5 pixels 
between each exposure.  These were later
merged to provide a properly-sampled spectrum with two pixels per
resolution element, and pixel size of
$6.5245\times 10^{-4}$~$\mu$m in the dispersion direction.
Each individual exposure lasted 60~seconds.
For each object exposure, an exposure of the nearby sky was also taken,
to subtract the strong OH night sky lines.
Our final co-added spectrum represents approximately 200 minutes on-source 
integration over the two half-nights of observing.

Our slit was placed at position angle 110$^\circ$ E of N, the same as
one of the observations in \citet{gon95}.  Our
slit width was essentially the same as theirs.  With
the one assumption that the K-band peak and the optical peak are
spatially coincident, we are observing almost exactly the same
region of the galaxy.

IRAF and routines we wrote for IRAF were used for basic
data reduction (flat-fielding, sky subtraction, combining undersampled 
spectra, wavelength and flux calibration, and removing the 
substantial tilt of the
spectral lines caused by a failure in the slit mechanism).  Careful attention
was paid to removing transient ``hot'' pixels, which were present in
the data.  We also carefully
corrected for small DC offsets in the wavelength scale apparent over
long timespans, probably caused by mechanical flexing.
After a thorough
examination, we believe we have properly accounted for all the instrumental
effects. 

OH night sky lines were used for wavelength calibration,
using the wavelengths tabulated by \citet{oli92}.
Because of the inevitable smearing induced by the fractional-pixel shifts
needed to remove the flexing,
we determined the actual spectral resolution of the data by measuring 
the OH line profiles in a co-added,
pure-sky image produced from the sky images obtained between the 
object exposures, and processed in the same fashion. 
The resolving power, found from the full-width at half maximum of the 
OH lines, was
$\sim$1200 (250~\kms) at 2.2~$\mu$m.

Spectra of A-type stars were used to remove atmospheric features and
correct for instrumental response, since
we could correct the only strong stellar feature expected, the Br$\gamma$ 
line, by fitting the line with a Gaussian and removing
it from the stellar spectrum.
The weather was not photometric on July 26, 1996, due to cirrus.
To correct for this, the spectra were scaled to a
common mean before co-adding them.  
We obtained absolute calibration from
the 1996 July 28 data, when the weather was 
better.  Observations of three flux standards showed the flux calibration 
to be accurate to $\sim$10\% in extractions 5 pixels wide in the spatial
direction.

The final product was a 2-dimensional flux spectrum with useful dimensions
of 500 (wavelength) by 49 (spatial) pixels, linearized in wavelength, 
and a corresponding
error spectrum.  Individual rows were then written out as ASCII files 
for further analysis using a stand-alone spectroscopic 
data reduction program described in \citet{gol95}.

\subsection{ JHK$n$ imaging}
\label{JHKimg.obs.sec}

We obtained near-IR  JHK$n$ images of NGC~7714 using the CASPIR
camera on the Australian National University 2.3m telescope at 
Siding Springs Observatory on the nights of October 30 and 31, 1994.
The K$n$ filter is centered at 2.165~$\mu$m with full-width at
half-maximum (FWHM) 0.33~$\mu$m.
The camera was based on a NICMOS3 256~$\times$~256 pixel
detector, with pixel size 0.5$\arcsec$.  The seeing, measured from
pointlike objects in the final galaxy images, was $\sim$1.6$\arcsec$
FWHM.  The central 1$'$ of our  H-band image 
is shown in Fig.~\ref{caspir}; the 
J and K$n$ images are virtually identical in appearance.

The images were taken in photometric conditions, and two sets of observations
were taken for each of two standard stars.  When measured through apertures
large enough to get almost all the flux from the stars (12$\arcsec$ diameter),
the stellar photometry agrees to 0.02 magnitudes.  Aperture corrections
are significant for the smallest apertures we considered.
Our photometry (with the flux standard and galaxy observed through
identical apertures) is given in Table\,\ref{NIRphot.tab}, and typically
agrees at the 5~--~10\% level with values from the literature. Our 
J \& H magnitudes are effectively on the CIT system, but because of
color terms between the  K and K$n$ filters, our  K$n$ magnitudes
should be corrected upwards by $\sim$0.1 magnitudes 
to be on the CIT system.

\subsection{HST UV imaging}
\label{UVimg.obs.sec}
Our HST images of NGC~7714 were taken with the WFPC2
through the F380W filter.  We used four exposures, 2$\times$500~s and
2$\times$400~s.  Our final image (Fig.~\ref{uvwfpc2a}) 
is the combination of the four exposures as calibrated
by the pipeline.  The filter was chosen as a compromise, since we 
were intending to image the sites of the
most massive stars, which are brightest in the UV, yet the sensitivity
of WFPC2 drops at the shorter wavelengths.  NGC~7714 was placed on the
PC chip. The full WFPC2 field of view was large enough to include the
nearby companion NGC~7715 as well. In Fig.~\ref{uvwfpc2a} we reproduce
the PC and WF frames at a cut to highlight the bridge between both
galaxies.

\section{Data analysis: the structure of NGC\,7714}
\label{data.anal.sec}

\subsection{The ultraviolet appearance}
\label{UVappear.sec}

In Fig.~\ref{uvwfpc2b} we show the central $2.5\arcsec \times 2.5\arcsec$ 
region of
NGC~7714 in the F380W filter, and, for comparison, in the F606W filter from
Paper~I. A spiral structure is seen, starting to the 
North of the nucleus and extending first to the East, then to the South-East
(where it reaches region\,A of Gonz\'alez-Delgado et al. (1995), 
about $2\arcsec$ off the plotted area). At least 20 luminous star clusters are 
visible, with the brightest cluster in the very center. Anticipating the 
discussion further below, we note that the nuclear region 
is less reddened than its surroundings.
Consequently, the brightness contrast between the nuclear and off-nuclear
clusters becomes less pronounced after obscuration effects are taken into
account. Surprisingly, the UV morphology of NGC~7714 from our new HST images is 
virtually identical to the morphology of the galaxy through
the red F606W filter (see Paper~I).  There are
dust lanes in and near the nucleus (Fig.~\ref{uvwfpc2b}), including
one which goes across the apparent geometric center of the nuclear
star-forming complex.  
The dark lanes are enhanced by making an ``unsharp-masked'' image,
subtracting an image median-filtered through a box 30 pixels wide.
We note that the brightest object in the nucleus, the cluster near
position (0,0) in Fig.~\ref{uvwfpc2b}, does not seem to be at
the geometric center of the star-forming complex, but appears to be located
$\sim0.3\arcsec$ NE of it.
We have made a ``color'' image by dividing the F606W by the F380W image.
The clusters are mostly very blue, and the dark lanes are redder,
further indicating that they may truly be dust lanes.

The F380W magnitude of the nucleus of NGC~7714 rises quickly with
aperture size up to aperture radii of about 2$\arcsec$, at which point the 
brightness increases much more slowly (Table\,\ref{Uphot.tab}).  As noted in 
Paper~I, this sets the spatial scale of the starburst region at a few 
hundred pc.

\subsection{The near-infrared appearance}
\label{NIRappear.sec}

In the infrared, NGC~7714 is dominated by a very small region centered
on the nucleus.  Nearly half the flux in a 9$\arcsec$ aperture comes from
the inner 1.74$\arcsec$ diameter region.
On larger scales, a bar is seen in the
IR images, at the same PA of $\sim$143$^\circ$ E of N as in optical images
\citep{gon95}.  The giant H~{\sc ii} region A is visible as a 
faint extension SE of the nucleus; region B is just
barely distinguishable from the spiral arm in which it is found, whereas it
is obvious in the HST images; and region C is readily visible 
SW of the nucleus. A higher-resolution (yet less deep) K-band
image was kindly provided to us by D.~Sanders. 
Taken in 1.1$\arcsec$ seeing (FWHM) with 0.37$\arcsec$ pixels, this image
shows that the nucleus is resolved and has a deconvolved size of 
about 1.3$\arcsec$ FWHM (with the FWHM of the
point-spread function removed in quadrature
from the FWHM of the nucleus in the image).
No spatial structure is apparent in the near-IR continuum 
of the nucleus itself, though we attempted
to deconvolve the image, and the nucleus may be slightly more extended in the
N~--~S direction. A similar result was found recently by Kotilainen et al.
(2000).

Red near-IR colors are often used as an indication of the presence of 
intermediate age populations (Persson et al. 1983, Silva \& Bothun 1998).
Near-IR colors derived from the data in Table\,\ref{NIRphot.tab} 
show a marginally significant radial gradient in $(J-Kn)$, the nuclear
values being redder than those of the surroundings. A more significant
gradient is found in $(H-Kn)$, while $(J-H)$ tends to be slightly
bluer in the nucleus.  Reddening is highly variable within the nucleus
even in the near-IR (Kotilainen et al. 2000).
After dereddening with $A_K \simeq 0.1$ (a typical value in
the successful models described further on) and converting $Kn$ to $K$,
we find $(J-K)_o \simeq 0.75$. This value is consistent with a
wide range of ages (e.g. Fig.\,2 of Lan\c{c}on 1999). The fact that
the spatial color gradient is stronger in $(H-K)$ than in $(J-H)$ 
may be taken as a case in favour of an intermediate age 
nuclear component, as cool AGB stars preferentially contaminate
the K band. However, 
anticipating the further discussion again, we note that colors
consistent with the data can be obtained with a variety of 
combinations of stellar ages and extinction patterns. The observed
gradient by itself cannot be simply transformed into a quantitative contribution
of any particular subcategory of cool stars.

\subsection{The infrared spectral image of the galaxy}
\label{NIRspec.sec}

A ``central'' spectrum of NGC~7714 was summed in a region
1$~\times$~5~pixels ($1.22\arcsec \times 6.10\arcsec$) in size, and is given in
Fig.~\ref{centspec}. The continuum resembles that
of a late-type (K/M) star, complete with the strong
CO absorption bands characteristic of such stars.  Atop this are
seen \Bg\ ($\lambda$2.166~$\mu$m), He~{\sc i} 
($\lambda$2.058~$\mu$m), and several 
H$_2$ emission lines.
The line and continuum fluxes are strongest in the few pixels centered 
on the nucleus.  However, continuum flux is also
detected well away from the nucleus, along the ``ring'' SE of the nucleus,
and also towards the NW.  
The \Bg\ and He~{\sc i} lines are visible as the slit crosses
the giant H~{\sc ii} region B, NW of the nucleus.
We also extracted the two strongest rows as the
``nuclear'' spectrum, which we will use later in conjunction with the
optical and UV spectra (Fig.~\ref{nucspec}).

The fluxes and equivalent widths of the three brightest lines as functions
of position along the slit are shown in Fig.~\ref{lines}.  As can be seen,
the equivalent widths of the ionized gas lines peak well off the nucleus, 
in H~{\sc ii} region B.
Interestingly, region B seems to have no significant IR continuum
emission of its own --- it is almost entirely an emission-line object.
This is consistent with its prominent appearance in the HST F606W
image (which includes the [O~{\sc iii}], H$\alpha$, and [N~{\sc ii}] lines), 
yet inconspicuous IR appearance. 
The lower equivalent widths in the central region are evidence that the
stellar flux is more concentrated towards the nucleus than the nebular
flux. Emission line strengths are given in Table\,\ref{NIRlines.tab}.

In Fig.~\ref{lines} we also show the profile of our spectral image 
along the spatial direction.  The profile was obtained by summing 55 pixels
in the dispersion direction, between the 2.12~$\mu$m H$_2$ line and
Br$\gamma$ line, for each spatial row of the spectrum.
The FWHM of the nucleus is about 3 pixels 
(3.6$\arcsec$).  Spectral images of the flux standards showed 
typical spatial FWHM of 2~--~2.5 
pixels (2.4$\arcsec$~--~3$\arcsec$) some of which was due to seeing, and some 
due to inaccuracies in centering the stellar image along the rows.
This implies that the seeing-deconvolved FWHM of the galaxy is of order
2.2~--~1.7 pixels (2.6$\arcsec$~--~2.0$\arcsec$).  This is larger than the 
deconvolved size as seen in the IR images, probably from a combination of
two effects.  First, the long (60~seconds) galaxy exposures are subject
to poorer seeing and greater image drifts than the 0.3~--~1~second stellar
spectrum images.  Second,
the H~{\sc ii} region A of \citet{gon95} is present in our slit just SE of the
nucleus, and this could also cause a wing
on the spatial profile of the nucleus.

Numerous stellar absorption features are observed, the strongest being
the ubiquitous CO band heads beginning at 2.3~$\mu$m.
The strength of 
the CO index in the ``central'' spectrum corresponds to a photometric
CO index of CO$_{\rm{ph}}=0.13$ magnitudes \citep{fro78}, and 
a spectroscopic CO
index of CO$_{\rm{sp}}=0.17$ magnitudes \citep{doy94}.  These values
are typical of K0-K2 supergiants or late-K to early-M giant stars.
In the ``nuclear'' spectrum, the CO index is CO$_{\rm{ph}}=0.19$
(CO$_{\rm{sp}}=0.26$), stronger than in the larger aperture, and implying the
presence of cooler and/or more luminous stars.  This is
evidence for a gradient in the stellar population in the inner few hundred
pc of the nucleus.

\subsubsection{Near-IR rotation curves}
\label{NIRrot.sec}

The observed velocities along the slit as determined from the \Bg, 
He~I, and H$_2$ $v=1-0$~S(1) lines are shown in
Fig.~\ref{kinematics}. The H$\alpha$ rotation
curve from \citet{gon95} is shown as a solid line.  
The agreement between the IR lines and the H$\alpha$ curve
is excellent SE of the nucleus.  To the NW of the nucleus, the \Bg\ line is
systematically below the optical rotation curve.
The discrepancy with \Bg\ might be due to the 
line profile being affected by a strong telluric line centered about 
1~pixel redward of \Bg, as the \Bg\ line
is redshifted into the night sky line.
The gaseous velocity dispersions were too small to be derived
at the spectral resolution of our data.

\subsubsection{H$_2$ gas in and around the nucleus}
\label{H2gas.sec}

An old question about the strong H$_2$ lines visible in the spectra of
starburst galaxies is whether the gas is excited by shocks, or by the
absorption of UV photons from the starburst \citep{gol97}.  
The populations of the different energy levels of the
H$_2$ molecules, reflected by the relative strengths of the different
emission lines, can tell if the gas was excited by a non-thermal mechanism,
such as shocks, or by the absorption of energetic photons.  For comparison
with our data, we refer to models S2 (for shock-excited gas, appropriate
at high H$_2$ densities) and 14 (for lines arising from UV fluoresence
in low-density gas) from \citet{bla87}.

In Table\,\ref{NIRlines.tab}, 
we give the measured strengths and upper limits (3$\sigma$, 
measured over a window 4~pixels wide in the dispersion direction) of
the H$_2$ lines in the ``nuclear'' and ``central'' spectra.  
To examine the gas farther from the nucleus, where different physical
conditions might prevail,
we extracted the spectrum of the brightest three pixels along the slit,
and subtracted this from our ``central'' spectrum.
This gave us the sum of two pixels with significant signal in them, but 
not including the peak of the nuclear starburst.  Though most of the lines
in the different spectra were not detected, the ratios of the
lines which were in fact seen 
(Table\,\ref{H2lines.tab}) are clearly different than for 
shock-excited, dense gas.
The data are consistent with a mix of both shock-excited and
UV-excited H$_2$.
This indicates that the total H$_2$ emission is dominated by
the UV-excited gas, because that emits a smaller fraction of its total
energy in the K-band lines than shock-excited gas.

\subsection{Summary of the central morphology}
\label{gradients.sec}

At UV, optical and near-IR wavelengths, NGC\,7714 has a rather
compact core. The $\sim 2\arcsec$ ``nucleus" that we focus on in this
series of papers, emits about half of the light 
measured within a $\sim 10\arcsec$ radius. Another factor $\leq 2$ is 
gained when including the whole galaxy.

While the near-IR continuum image is smooth in the nucleus, the UV
and optical images show complex structure. It is instructive to 
compare them with the radio maps of Condon et al. (1982) and 
the 2.16\,$\mu$m Br$\gamma$ emission map recently obtained by Kotilainen et al.
(2000). Both show a double peaked morphology within our
``nucleus", with two maxima
separated by about $1\arcsec$ along a NE-SW line. The best correspondance
between all maps is obtained if region~2 of
Kotilainen et al. (2000) is associated with the extended UV source at
$\sim 2.2\arcsec$ to the East of the geometrical center, at the very
edge of our images in Fig.\,\ref{uvwfpc2b}. The brightest $2\arcsec$ in
the near-IR continuum or in the UV then cover most of 
regions 1 and N of Kotilainen et al. (2000). With this assumption,
the Br$\gamma$ and H$_2$ fluxes of these authors and ours are
in excellent agreement. The brightest spot in the Br\,$\gamma$ map
does not coincide with the bluest, brightest UV cluster. It is associated
with less conspicuous UV sources, to the SW of the brightest one.

The evolution of the UV to near-IR flux ratio with aperture size was
reconstructed from the magnitudes in Tables \ref{Uphot.tab} and 
\ref{NIRphot.tab}. It is shown in Fig.\,\ref{aperture_col_lin}. 
$(F380W-Kn) \simeq 3.06$ in the central $12\arcsec$. 
Remembering the 0.1 magnitude
difference between $K$ and $Kn$, this agrees well with 
previous measurements. $(U-K) \simeq 2.9$ is obtained from the data
summarized by Smith \& Wallin (1992). The UV/near-IR flux ratio
decreases steadily from $5\arcsec$ diameter apertures to larger ones,
confirming the suspected trend of increasing stellar age or extinction
with radius. Relatively red colors are obtained from our data
for the ``nucleus". Although this 0.15\,magnitude effect may
be partly due to aperture mismatch, we estimate that it is 
significant. However, in view of the preceding discussion, it
should not be interpreted as a simple gradient, but as the result of a
patchy distribution of the luminous hot stars and the dust.

\section{Understanding the composite nuclear spectrum of NGC 7714}
\label{Spectrum.sec}

Our efforts to fit the nuclear spectrum of NGC~7714 are based on
a series of models from P\'egase \citep{fio97} and the Starburst99
project \citep{lei99}.
Starburst99 is an update of the work of \citet{lei95}. The reader
is the refered to the three quoted papers for details. Both P\'egase
and Starburst99 utilize an evolutionary synthesis
technique, with the former emphasizing the physics of cool stars
(in particular the inclusion of the thermally pulsing AGB; Lan\c{c}on
et al. 1999), and the latter that of hot stars. In both approaches,
models of the emission from starburst stellar populations are
calculated as functions of age, the initial mass function slopes,
lower and upper mass cutoffs,
and stellar evolutionary tracks of selected metallicities.
 Continuum emission from ionized gas is always taken into account, with
the assumption that no Lyman continuum photons escape from the
galaxy. As all stellar populations considered will have total stellar masses
larger than $10^6$\,\Ms, the effects of stochastic fluctuations in the
numbers of luminous stars can be neglected (Lan\c{c}on \& Mouhcine 2000,
Cervi\~no et al. 2000). Unless otherwise stated, a Salpeter IMF extending from
1 to 80\,\Ms\ is used as it was done in Paper~I.

We attempt to arrive at the simplest possible models explaining the
spectrum of the very center of NGC 7714.  As we will discuss,
model solutions are not unique. In order to further restrict the 
range of satisfactory models, the UV to near-IR SED is  
combined  with information from longer and shorter wavelengths.
We aim at identifying the common features of the collection of
simple models that are consistent with the data.

\subsection{The composite nuclear spectrum}
\label{composite.sec}

We have used our data to construct the multiwavelength SED of 
the inner $\sim$330~--~380~pc ($1.74\arcsec - 2\arcsec$) of NGC~7714.
For the UV, we have used the GHRS spectrum of Paper I, which 
was obtained through the $1.74\arcsec \times 1.74\arcsec$ Large Science
Aperture.  For the optical spectroscopy, we averaged two spectra 
extracted by \citet{gon95},
one in aperture $1.2\arcsec \times 2.1\arcsec$, and 
the other in $1.2\arcsec \times 3.5\arcsec$.
This gives us an effective aperture a bit larger than the GHRS aperture.
We scaled up the optical spectrum fluxes by a factor of 1.6 to account
for point-source losses through the narrow
slit (see the note in Paper~I).   We performed photometry on our
HST images through apertures 1.74$\arcsec$ in diameter.
We used the ``nuclear''  K-band spectrum in an aperture 
$1.2\arcsec \times 2.4\arcsec$.  
From our IR images, we obtained photometry at JHK$n$ through circular
apertures 1.74$\arcsec$ in diameter, to approximate the GHRS aperture.

Though we have tried to obtain data through similarly-sized apertures, it
is important to note that slight differences in aperture position, size, 
and photometric calibration must cause some normalization differences
between the spectra.  We estimate those differences are at about the
10~--~15\% level at most.

\subsection{Model boundary conditions}
\label{constraints.sec}

Several basic constraints must be considered while fitting the 
multiwavelength nuclear spectrum of NGC 7714.  These contraints result 
from our previous studies (Paper\,I), from the literature
and from the above data analysis. 

\begin{enumerate}
\item The nuclear chemical composition is close to solar.
Our models are made under this assumption (but see Sect.\,\ref{consistency.sec}
for a discussion).

\item The observed UV spectrum in the GHRS aperture is best 
described as an instantaneous burst population of age $\sim$4.5~Myr
with reddening of $E(B-V)\approx 0.11$.  Of this reddening, about 0.08 
magnitude is from the Milky Way, and 0.03 magnitude 
is intrinsic to NGC~7714.

\item The reddening to the nuclear ionized gas, determined from the 
Balmer decrement, is $E(B-V)=0.21$.  About 0.13 magnitude is intrinsic to 
NGC~7714. 

\item A strong 4000~\AA\ break is present in the optical spectrum, implying 
that a significant part of the optical spectrum is due to a population
older than $\sim 10^8$\,years. 

\item A luminous near-IR continuum is present,
and the conspicuous CO bands at
2.3\micron\ show that this continuum is due to luminous cool stars.

\item The IRAS far-IR luminosity is $L_{\rm{IR}} =  3 \times 10^{10}$~\Ls.
$L_{\rm{IR}}$ approximates the total bolometric luminosity and exceeds 
the UV luminosity in the IUE aperture by about an order of magnitude 
\citep{hec98}. The UV
luminosity in the GHRS aperture is another factor of $\sim$2 smaller
(Paper~I). One may a priori expect the nuclear stars to be responsible for
$10\,\% - 50\,\%$ of the IRAS flux. Noting that no direct spatial
information is available at the longest IR wavelengths, this
is consistent with the recent 12$\mu$m maps obtained with the Infrared 
Space Observatory (O'Halloran et al. 2000). 

\item The ROSAT soft X-ray luminosity of NGC~7714 is 
$4.4 \times 10^{40}$~erg~s$^{-1}$. \citet{ste98} interpret this X-ray
flux as due to thermal emission from supernova powered, hot bubbles and/or
a galactic wind. The non-thermal 20~cm flux reported by \citet{wee81} is
most likely related to the same phenomenon. 

\end{enumerate}

Although our goal is to fit a co-spatial region in the nucleus
of NGC~7714, we recall the properties
of the spectrum measured in the IUE aperture. That spectrum suggests a
population of OB stars, possibly mixed with some older stars.
Typical values for the galaxy's $E(B-V)$ are around 0.3.

\subsection{Dust obscuration and emission}
\label{dust.sec}

The far-IR emission of NGC\,7714 shows that dust cannot be neglected
in the study of the energy distribution of the galaxy.
The presence of dust will affect the shapes of modeled spectral energy
distributions, in a way that depends both on the nature of the
grains and on their spatial distribution relative to the sources of
light. 
\citet{pux94} used optical and near-IR recombination lines to examine the
dust obscuration towards ionized gas in NGC\,7714. They found that the line
ratios were adequately fitted by a simple foreground screen of
dust with $A_{\rm{V}}=0.86 \pm 0.13$ magnitudes
($E(B-V)=0.28\pm0.04$). Since the simplest possible extinction geometry
fitted their observations quite well, they felt no need to favor the
more complex alternatives they had explored. 

In our attempts to fit the SED of NGC~7714, we have systematically
accounted for the 0.08 magnitudes of foreground Galactic extinction using a 
standard extinction curve (Howarth  1983).  For extinction intrinsic to
NGC 7714, we generally use the starburst obscuration curve of \citet{cal00}, 
in the framework of the foreground screen geometry consistent with
the derivation of that curve. This seems reasonable in the
light of the results of \citet{pux94}, and avoids introducing more free 
parameters associated with more complicated extinction geometries.

We do however also invoke different choices of the geometry
and the extinction law. Spatial variations of the extinction are
to be expected in NGC\,7714: dust lanes cross the images, extinction
in the GHRS and IUE apertures differ. 
The obscuration curve of Calzetti et al. (2000) was derived 
empirically from a sample of integrated spectra of starburst galaxies,
in order to provide a tool to be used in a screen geometry formalism 
for average starbursts. It implicitly accounts for both wavelength 
dependent dust properties and typical dust distributions. 
The distribution appropriate for the central few arcseconds of NGC\,7714 may 
differ from this average picture. 

Our predictions for the far-IR emission are based on a simple energy budget.
We compute the difference between the model SEDs before and after extinction
has been applied, integrate the difference over wavelengths and assume that 
this energy is reradiated by the dust in the far-IR. We note that 
this method assumes symmetry in the galaxy structure. For instance, it
underestimates the far-IR flux if UV-optical light escapes towards
the observer more easily than in any other direction from the source,
and it overestimates the far-IR flux if the line of sight happens to
cross a particularly dense foreground cloud. The wavelength
dependence  of extinction ensures that the hottest of the reddened stars
contribute most strongly to the predicted emission. The emission temperature
of the dust is known to depend on the grain size distribution and on 
the proximity of the grains to the hot stars
(Helou 1986, Mazzarella et al. 1991). This also determines which 
fraction of the dust radiation was actually measured by IRAS. However,
the absence of spatial resolution in the far-IR already results in such
a large uncertainty on the nuclear emission that it is not justified to
attempt to model the far-IR energy distribution more precisely.

\subsection{The nucleus approached from three wavelength ranges}

\subsubsection{Results from the GHRS UV spectrum}
\label{UVresults.sec}

Paper~I showed that a 5\,Myr old instantaneous burst (ISB) 
with an intrinsic $E(B-V)$ of 0.08 provides
the best representation of the stellar lines seen in the GHRS UV
spectrum. The corresponding reddened model spectrum, scaled
to match the GHRS spectrum, accounts for about
$30 \pm 3$\% of the continuum flux at 5500~\AA\ 
and only about 3\% of the K-band continuum.  This population
produces about 1\,\% of the galaxy's far-IR luminosity if the
dust optical depths around it does not exceed the optical
depth on our line of sight. In the more likely case of an irregular dust
distribution, the burst would provide $2-4\,\%$ of the far-IR
light ($4\,\%$ corresponds to the bolometric luminosity of this burst
population).

The percentage contribution of the 5\,Myr burst at near-IR wavelengths 
can be increased with dust distributions more complex than a screen. 
But no reasonable adjustment of the UV to near-IR SED can be obtained with 
this single burst age. Such adjustments would require more than
99\,\% of the young stars to be heavily obscured, leading to unrealistic 
amounts of far-IR emission (besides being incompatible with the spectroscopic 
information in the optical and near-IR). 
Thus, the young burst seen in the 
UV does not by itself explain the energy distribution of NGC\,7714's nucleus. 

The UV spectrum is marginally consistent with models for ISBs 
slightly older than 5\,Myr or with continuous, constant star formation (CSF). 
Again, none of these, combined with a simple foreground screen extinction,
can fit the observed colors. Adjustments of the broad band energy
distribution with CSF models and modified dust properties
can be obtained, but since they do not follow directly from the UV analysis
their discussion is postponed to Sect.\,\ref{onepop.sec}.


The conclusion from the UV analysis, placed in the context of the 
UV-through-IR SED of  NGC\,7714, is that the recent
star formation event, though dominant at short wavelengths, is
not the only relevant star-formation episode to be considered, even
when focusing on the central 330~pc of the galaxy. 
What is most likely required is one population of age $\sim$5~Myr, and
at least one older population rich in cool stars and weak in ionizing flux.

\subsubsection{Results from the Balmer line profiles and the Balmer jump}
\label{BVresults.sec}

The optical spectra of starbursts 
are dominated by nebular emission lines, but they
can also display the H and He~I absorption lines formed in the photospheres 
of O, B, and A stars. The reason why these features can indeed be detected 
in absorption in starbursts such as NGC\,7714
is that the equivalent width of this absorption is constant
throughout the Balmer series, while the emission line strengths
decrease rapidly with decreasing wavelength.
The strengths of the Balmer and He~I absorption 
lines show a strong dependence on effective temperature and 
gravity, and therefore provide constraints on starburst and 
post-starburst ages (Gonz\'alez Delgado \& Leitherer 1999,
Gonz\'alez Delgado et al. 1999). In the spectrum of a coeval population, 
the absorption equivalent widths of theses lines increase 
with time until they reach a maximum at ages of 300~--~500~Myr, and decrease
afterwards. Spectra for older and younger populations can
also be distinguished through the shape of the ``continuum'' 
in the Balmer jump region, between 3720 
and 3920\,\AA: the rather abrupt jump seen shortward of 3800\,\AA\ 
at young ages turns into a flatter but broader slope later on 
(e.g. Fig.\,2 of Gonz\'alez Delgado et al. 1997).

For an instantaneous young burst (a few Myr old), the
predicted Balmer lines are much weaker than those observed in the
nucleus of NGC\,7714. 
This confirms the need for a composite nuclear stellar population. 

A variety of star formation histories are consistent with the
absorption line profiles of the normalized optical spectrum of the
nucleus of NGC\,7714. Among them, a CSF model with a duration of a few 100\,Myr,
or combinations of a 4~--~5~Myr old burst with 
populations a few 100~Myr old. However, some of these fail to simultaneously
reproduce the shape of the ``continuum" in the Balmer jump region.
There is no abrupt Balmer jump in the NGC\,7714 spectrum. In
the satisfactory models, the relative fraction of young ($<100$\,Myr old), 
hot stars must be smaller than in constant star formation models,
unless star formation lasts for more than $\sim$800\,Myr.

\subsubsection{Results from the near-IR spectrum}
\label{NIRresults.sec}

As mentioned in Section\,\ref{NIRspec.sec}, the near-IR spectrum 
shows numerous stellar absorption features. The CO bands are 
relatively strong, implying the presence of red supergiant stars,
but not strong enough to exclude a significant contribution from
less luminous late-type giants. These can be predominantly upper AGB 
stars if the typical ages lie between 10$^8$ and $10^9$\,yr, 
or RGB stars for older ages. 

After a few 100\,Myr of evolution, about 50\,\% of the K band light of an 
instantaneous burst population originates from upper AGB stars. The
light thus carries the spectral signatures of long-period variables that may
be oxygen-rich (e.g. Miras) or carbon-rich \citep{lan99}. 
An inspection of the stellar spectra of \citet{lawo00}
shows that the metal lines identified in the nucleus of NGC\,7714 are
present with various strengths in oxygen-rich late-type stars of all
luminosity classes, including Miras. In contrast, they
are not seen in C stars, where blends of many other features mainly due to
C$_2$ and CN make the K band pseudo-continuum much more irregular
(at the resolution of our data). 
 None of the typical C star features is obvious in the 
NGC\,7714 spectrum (Fig.\,\ref{N7714_IR_stars.fig}, top). 
Thus, although C stars can be more numerous than
Miras in particular circumstances (e.g., at low metallicity and a 
metallicity-dependent 
range of ages; Iben \& Renzini 1983; Groenewegen \& de Jong 1993;
Mouhcine \& Lan\c{c}on, in preparation), we do not need to take this
possibility into account here.  Many variable O-rich AGB stars show
strong H$_2$O absorption bands around 1.9\,$\mu$m, whose wings extend  
up to 2.1$\,\mu$m. The resulting change in the apparent continuum slope
is not seen in our NGC\,7714 spectrum
(Fig.\,\ref{N7714_IR_stars.fig}, second from top). However, we note that
at wavelengths below 2.1$\,\mu$m the correction for telluric H$_2$O absorption 
is important, possibly affecting the flux levels with larger errors.
In addition, 
because of uncertainties in the evolutionary tracks and the effective 
temperature scales of the upper AGB, the quantitative predictions of the 
synthesis models regarding the strength of the water bands are still uncertain.

In summary, the K band spectrum is consistent with a variety of combinations 
of red supergiants, AGB stars and RGB stars. It is unlikely that a
single red supergiant dominated population (e.g., an ISB with an age
of a few 10$^7$\,yr) or a single AGB dominated population (e.g., an
ISB with an age of several 10$^8$\,yr) can quasi-exclusively be
responsible for the near-IR light. A similarly open-ended conclusion was
drawn from a near-IR spectrum at shorter wavelengths by
\citet{gon95}, on the basis of the calcium
triplet absorption lines.

\subsection{Simple star formation histories that adjust the broad band
SED}
\label{onepop.sec}

{\em Pluralitas non est ponenda sine necessitate} (Occam's Razor) ---
yet a single coeval young population has been excluded
as a representation of the nucleus of NGC\,7714
in the previous sections. The next, more complex model has
constant star formation over some period of time.

\citet{gor97} studied the starburst in NGC\,7714 (through
larger apertures than ours) using a Monte-Carlo radiative
transfer model with different dust distributions relative to the
stars. They found that, among others, a long lasting CSF model with a 
duration of about 1\,Gyr could represent the data, 
assuming SMC dust properties and an inhomogeneous shell geometry
(dust shell with higher density clumps). 
Our aim is not to explore all possible dust geometries. 
However, the degeneracy 
between intrinsic SED properties and obscuration properties mentioned 
by \citet{gor97} cannot be ignored. It is indeed
possible to fit the broad band energy distribution even of the 
central 2$\arcsec$ with constantly star-forming models
when the dust distribution is allowed to differ from a simple
screen. The HST images and color maps clearly allow for some
complexity. For instance, the SED is reproduced very well after 
1\,Gyr of constant star formation, if nearly half the stars 
are hardly reddened at all and the other half are located behind
dust screens with optical depths of the order of 1.5 
(Figure\,\ref{1pop.2screens.fig}).
This configuration predicts that $4\times10^9$~\Ls\ of stellar light
is absorbed by dust, with a predominant contribution from the
hotter of the coexisting stars; the nuclear 5$\arcsec$ then provide about 
10\,\% of the total far-IR luminosity of the galaxy, a fraction
consistent with the galaxy morphology as summarized in 
Sect.\,\ref{constraints.sec}. 

How much can the age of the CSF model be varied, if only dust
is allowed to modify the SED in order to match the broad band photometry? 
For CSF durations larger than 5\,Gyr, the intrinsic $(UV-V)$ color 
of the stellar population becomes redder than observed in NGC\,7714. 
Older models cannot be considered (scattering may
reduce the reddening effect of dust, but ``blueing" in the
optical is not expected; Witt et al. 1992). Already
at 5\,Gyr, the contribution of the reddened component is so low that
only 2\,\% of the far-IR light of the galaxy are accounted 
for in the nucleus. Going towards lower ages, CSF models as young
as a few 10$^7$\,yr reproduce the observed colors, with adequate combinations
of dust optical depths. However, the large number of hot obscured
stars then tends to produce too much far-IR flux. Both the old and young 
extreme CSF histories considered also produce inadequate spectral
signatures in the Balmer region.
If star formation is assumed to have been constant in
the nuclear area but dust reddening is allowed to vary with
the line of sight, ages between about 800\,Myr and about 3\,Gyr
match the colors and the optical absorption features, and are
also consistent with the near-IR spectrum.

The stellar features observed in the GHRS UV spectrum, however, are so strong
that consistency with a long episode of constant star formation is only
marginal. A larger relative UV contribution of stellar populations with an age 
of the order of 5 Myr is favored. Model options based on this constraint from 
Paper~I are explored in the following section.

\subsection{Adjusting both the colors and the spectra}
\label{multipop.sec}

As soon as we stop limiting ourselves to single ISB or CSF models
but consider composite populations, the number of
model parameters increases dramatically. 
Throughout the following, we maintain the constraint
that the UV continuum of the GHRS aperture predominantly arises
from a 4.5~--~5 Myr old instantaneous burst with very low intrinsic reddening.
A ``component" of the composite population is defined as group
of stars that can be represented with a common, simple star formation
history: an ISB, a CSF, or an exponentially decreasing star formation
rate. For simplicity, a same amount of obscuration is assumed to affect 
all stars of a same component. 

In practice, we proceed the following way. Various families of models 
are considered successively, defined by the number of stellar populations
considered in addition to the GHRS burst (limited to 2 or 3), the type of 
star formation history in each (e.g. ISB, CSF, exponential), and
the adopted extinction prescription. For each family, 
we step through component ages. An automatic optimization procedure (based on
a quality-weighted $\chi^2$) then determines the most adequate extinction
values and relative contributions for the components. By visual
inspection, we reject solutions that are obviously inconsistent 
with the data; the others are included in the discussion. The choice
of a soft rejection criterion is dictated by the uncertainties in
the observations, and in particular in the relative calibration
of the individual spectral segments.

In order to identify global properties of the missing stellar components,
we first consider models with either {\em one} additional burst or 
{\em one} additional constantly star-forming component. For a given age of the 
second component, both its relative contribution to the light and the amount 
of reddening it is affected by are obtained by optimizing the fit to 
the UV through near-IR spectrum. The best fits obtained with the assumption of
constant star formation qualify as good, while the additional burst
assumption merely provides a marginally acceptable fit, in particular
around 4000\,\AA\ and for the CO bands. Both types of models provide
satisfactory amounts of far-IR emission. The two classes of results are shown 
schematically in Fig.\,\ref{2pop.schematic.fig}\,: either a massive
starburst occurred about 300\,Myr ago, or star formation occurred
continuously over about 800\,Myr at a rate of about 0.6~\Myr. 

Despite their differences, the two classes of selected ``two-component" models 
have an important common feature\,: over timescales of a few to several
10$^8$\,yr, $5-7\times 10^8$~\Ms\ of stars were formed in the
observed aperture (in the $1-80$~\Ms\ mass range). The presence
of more stars in the age range of $10^7-10^8$~yr is, in the constant star 
formation scenario, compensated for with an older age, in order to adjust
the spectrum around the Balmer jump. The $5\times 10^6$ to $10^7$~\Ms\ 
of stars formed over the last 5\,Myr in both cases are only a small part of 
a more important star formation episode.

What, then, was the history of  star formation over the last several 
10$^8$~yr?

A simple two-burst model is most probably inadequate.
In addition to providing a marginally acceptable fit to the spectrum, 
it requires fine tuning to explain the observed radio emission\,: 
the supernova rate 
predicted is too low because the 5\,Myr old population is still too young
and the 300\,Myr old burst is too old. As seen in the HST images, 
compact very blue knots are found at several locations. No obvious
feature of the galaxy explains why they should have formed simultaneously
5\,Myr ago after a long period of quiescence, whereas they 
can quite naturally be seen as the youngest components of a more
extended process of star or star cluster formation, for instance as a result
of the gravitational interaction with NGC\,7715. 

On the other hand, a timescale of 800\,Myr is long compared to estimates
of the interaction timescale with NGC\,7715 (Smith et al. 1997). 
Shortening the star formation timescale without producing too many ionizing
photons and degrading the fit around the Balmer jump, requires the partial 
suppression of very recent star formation (Sect.\,\ref{BVresults.sec}). 
In this context, the most natural explanation for the observed
properties of the nucleus of NGC\,7714 is a globally 
decreasing star formation rate with an initiation 
a few 10$^8$~yrs ago. The data are consistent with a smooth, 
exponentially decreasing star-formation rate with an exponential timescale 
of 50~--~100~Myr, or with a series of burst events (as little 
as three may be sufficient) of decreasing total masses. 
In the latter case, a 15~--~50\,Myr old burst provides the 
required supernova rate, a significant contribution of red supergiants to the 
near-IR flux, and usually a dominant fraction of the far-IR luminosity; the 
intermediate age burst (a few 100\,Myr old) adjusts the spectrum in the 
4000\,\AA\ region, contributes significantly to the near-IR light
because of upper AGB stars, and participates at a lower but 
non-negligible level to the heating of dust. Examples are shown
schematically in Fig.\,\ref{3pop.schematic.fig}, and an illustrative
fit is plotted in Fig.\,\ref{3pop.fit.fig}. 

Finally, a contribution of a pre-existing, old population may be expected
in the nucleus of a spiral galaxy. 
The amount of this contribution depends strongly on the model adopted 
for the underlying spiral. And very little is known a priori on the relative
enhancement in the mass of stars a starburst can produce. The spectrum
itself provides constraints. Old populations have much redder colors
in the optical range than observed in the nucleus of NGC~7714. They
cannot be dominant; a strong contribution from an intermediate
age population (less than 1\,Gyr in age) remains necessary.
A large number of tests were performed, in which the star-formation 
history of the underlying spiral was represented with a Schmidt law, as 
appropriate for Sa-type galaxies. It is indeed possible to 
obtain satisfactory adjustments  of the nuclear data
with a significant contribution of the underlying population 
(e.g. Fig.\,\ref{3pop.underl.fig}). The duration of the recent 
star-formation episode can then be relatively short (down to about 200\,Myr), 
as the pre-existing stars help providing a good fit 
around 4000\,\AA. With star-formation durations larger than 200\,Myr,
however,
the upper AGB stars help in reaching the observed near-IR flux level and less
fine-tuning is required. 

The most important effect of the underlying population is to reduce the
mass of stars produced by the intermediate age events from typically
$3-4 \times 10^8$\,M$_{\odot}$ to about $1-2 \times 10^8$\,M$_{\odot}$.
The global star formation scheme however remains valid.
Most of the far-IR flux is due to the hot stars of the recent events, though a
(model dependent) contribution from the underlying population is 
present (current star formation is not nil in the underlying spiral
models). The mass of stars in the underlying population can 
exceed the mass in recent stars. It reaches about $10^9$~\Ms\ only in a 
rather extreme category of models where a very old, nearly constantly 
star-forming spiral population would totally dominate the near-IR emission. 
This would help explaining the very regular near-IR continuum
images of the nucleus, but is difficult to reconcile with the strength
of the observed near-IR absorption features. 
We conclude that the mass of stars formed over the last few $10^8$\,yr 
is at least 10\,\% of the total stellar mass present in 
the aperture, and we favor values around 25\,\%.

\subsection{Further consistency checks and plausible model variants}
\label{consistency.sec}

In the study described above, we have assumed that the stellar populations
of NGC\,7714 have solar metallicity. As the nebular lines suggest slightly 
subsolar abundances (Paper I), we have also systematically searched for models 
at half-solar metallicity. We note that these must be treated with
some caution for supergiant-dominated post-starbursts: current 
stellar evolution tracks produce a smaller number ratio of red to blue 
supergiants at lower metallicities, while star counts favor the
opposite trend (Langer \& Maeder 1995; Origlia et al. 1999).
A more robust prediction is that subsolar populations
have bluer red giant and asymptotic giant branches. 
A global consequence of the use of subsolar tracks is the need
for more reddening to adjust the observed SED. It becomes more 
difficult to explain the large near-IR flux unless some of the cool stellar 
populations are affected by as much as 1.5\,magnitudes of visual
extinction (screen model). However, albeit with differences in the 
precise values of the model parameters, we find that the range of
successful models at half-solar metallicity essentially overlaps
with those found with solar abundances. The predicted far-IR emission
is larger, but not enough to modify our conclusions. While the optical
emission line ratios at Z$_{\odot}$/2 clearly favor an instantaneous
5\,Myr old burst as the origin of the Lyman continuum photons (Paper I),
both ISB and CSF models predict ratios consistent with the data
at the solar metallicity used here.

A further check on the validity of our fits is how well we can reproduce
the observed hydrogen recombination line equivalent widths.
As an illustration, we will focus on the 3-burst model of
Fig.\,\ref{3pop.fit.fig} (we note that the presence of the
underlying population does not modify the reasoning and results).
The H$\beta$ equivalent width ($W(\rm{H}\beta)$) is measured
to be $\sim$26~\AA\ in the optical spectrum we used, and 
$W(\rm{Br}\gamma)$ is 21~\AA\ in our K-band spectrum.  In our
model, the UV-bright, 5~Myr population is responsible for about 25\% of
the continuum at H$\beta$, and 3\% of the continuum at Br$\gamma$.
The emission line fluxes of the 200~Myr population and the
10~Myr population are negligible.  The contributions of the other
stellar populations dilute the equivalent widths from the youngest
population to 20~\AA\ at H$\beta$ and 5~\AA\ at Br$\gamma$.  However,
the emission line equivalent widths change strongly between 3 and 5~Myr,
and if the age of the youngest component is taken as 3~Myr instead of
5~Myr, then the diluted equivalent widths become 55~\AA\ at H$\beta$ and
14~\AA\ at Br$\gamma$.  Hence, we can trade off small age differences
against fractional contributions of the youngest component at any 
particular wavelength.  We can account for the observed H$\beta$ equivalent
width this way, but the predicted Br$\gamma$ equivalent width is still 
too small by up to 50\%.

Extinction can solve this apparent weakness of the model
in various ways. It has often been stated that the gas emission 
associated with a starburst is reddened more than SED 
(Calzetti, 2000). If this is the case in the nucleus of NGC\,7714, 
the agreement with the ratio of the Br$\gamma$ to H$\beta$ equivalent 
widths is improved, but the deficiency of Br$\gamma$ photons remains.
More likely, 
there may be additional young stellar populations, obscured 
enough to not contribute significantly to the UV SED (a requirement
imposed by the blue UV colors). The reddened line emission 
associated with these stars enhances the equivalent widths of the Brackett 
lines more than those of the Balmer series. This configuration explains 
both the missing Br$\gamma$ flux and the fact that the line ratios lead to 
a higher $E(B-V)$ than the UV continuum. It is fully 
consistent with the nuclear morphology discussed in 
Sect.\,\ref{gradients.sec}. The hidden young populations must be able
to increase the Br$\gamma$ equivalent width by the necessary factor 
without degrading the fit to the SED (in particular in the UV and around 
the Balmer jump). We find that they may contain up to 4 times the stellar mass 
of the burst actually seen in the UV, and would be typically
be affected with $E(B-V) \simeq 0.35$. The modified models remain
consistent with the far-IR constraints ($10^{10}$\,L$_{\odot}$ of far-IR
light are predicted for the nucleus). We emphasize that the 
presence of the hidden stars does not drastically modify the 
scenarii we have already considered. The most recent star formation 
remains a small part of a longer star formation episode.
In some of the previously discussed successful models, the GHRS burst 
was combined with continuous star formation components: the latter 
can also provide the obscured hot stars.

In Paper~I, the bolometric luminosity of the starburst {\em seen} in the
UV continuum of the GHRS aperture was estimated as $5\times10^{9}$~\Ls, 
almost 10 times less than the IRAS luminosity. The range of models
suggested here agrees with this result. That young starburst 
population can at most be responsible for a few $10^9$\,\Ls\ 
of dust emission in the far-IR, the exact value depending on the 
distribution of dust around those stars, in directions other than
our unobscured line of sight. The hottest of the stars associated
with the intermediate age populations, which formed stars over the
last few 100\,Myr and are more obscured, typically account
for $3-6\times 10^9$\,\Ls\ of dust-processed far-IR light, i.e. 
less than 20\,\% of the IRAS luminosity. A contribution of
up to 50\,\% might be achieved without too much stretching if, 
as suggested by the emission line analysis, additional obscured young 
populations are present. In any case, a significant fraction of the far-IR 
luminosity is expected to come from the stellar populations distributed 
outside the nucleus. As the IUE flux at 1500~\AA\ is about twice the GHRS 
flux and the IUE continuum is globally affected by reddening, this is
a reasonable conclusion. Only high spatial resolution in the mid/far-infrared
would allow us to derive tighter constraints from the energy budget.

We showed in Paper~I that the 5~Myr burst seen in the UV
could account for the observed X-ray and non-thermal radio
emission of the galaxy if it produced supernovae and supernova
remnants which have unusual (but not implausible) properties. The
5~Myr burst is associated with a supernova rate of 0.007~yr$^{-1}$.
If the supernovae power the X-ray luminosity of
$6 \times 10^{40}$~erg~s$^{-1}$ and the non-thermal radio luminosity
of $4 \times 10^{38}$~erg~s$^{-1}$ \citep{wee81}, radiation losses
must be small and the supernova remnants must be excessively
luminous. An underlying obscured population of age tens of Myr
greatly alleviates these constraints. For instance, the
3-component model in Fig.~\ref{3pop.fit.fig} has a 20~Myr
old population producing supernovae at a rate of 0.07~yr$^{-1}$,
i.e. an order of magnitude larger than the supernova population
from the unobscured 5~Myr burst. This rate is very similar to
the one derived for the {\em IR-bright} starburst archetype M~82:
\citet{ulv94} find a supernova rate of $\sim$0.1~yr$^{-1}$ in M~82.
M~82 and NGC~7714 have the same bolometric luminosities of
$3 \times 10^{10}$~\Ls. If they are powered by starbursts, and if the
starburst properties are similar, the supernova rates should be
similar as well. The proximity of M~82 permits detailed radio mapping
and relatively accurate supernova and supernova remnant estimates.
NGC~7714 is too distant for such studies. The {\em measured} supernova rate
in M~82 and the {\em predicted} rate in NGC~7714 agree. Since the
predicted rate follows from our multi-component fitting of the
entire SED, we can have some confidence in the overall sanity of the
models.

\section{Discussion}
\label{Discussion.sec}

A wide range of models has been shown to agree
with the available information on the SED of the central $\sim$330\,pc of
NGC\,7714. Yet, the global scheme of all of them is similar,
and in good agreement with the current understanding of starburst galaxies\,:
star formation in a starburst partly occurs in compact brief bursts
and may globally be spread over regions of several 100\,pc in space and several
100\,Myr in time, mimicking a relatively continuous and extended
star-formation episode.

Smith \& Wallin (1992) and Smith et al. (1997) used dynamical simulations
to study the NGC\,7714/7715 system. They estimate that 100\,Myr have
elapsed since closest approach. As confirmed by J. Wallin (private
communication), the scaling uncertainties in the dynamical models
would not allow us to stretch the interaction timescale to much
more than 200\,Myr, which is among the very shortest of
the star formation timescales of satisfactory spectrophotometric models.
We interpret this as evidence that the latest interaction, which certainly
played an important role in triggering the most recent star formation, probably
followed other events that had already favored enhanced star forming activity
earlier on. A second pass of NGC\,7715 is one of the scenarii worth further
investigation in the future.

Which data are most important in constraining the many evolutionary
model parameters? A vast range of models was found
capable  of reproducing the {\em broad-band} energy distribution, in particular
because of the strong degeneracy between intrinsic colors
and reddening introduced when the dust distribution is allowed
to be complex (as suggested by the images). The parameter space
was further narrowed using (i) the stellar lines in the UV --- because
they were strong enough to imply a predominant 5\,Myr old population
at short wavelengths, (ii) the spectrum around the Balmer jump --- because the
absorption lines and the continuum are sensitive to age, and thus
to the relative contributions of various components, in the age range
relevant to NGC\,7714. The K band spectra, then, were used
for consistency checks: none of the specific spectral signatures
of red supergiants, AGB stars or red giants happen to be strong enough in
NGC\,7714 to imply the predominance of one of these types,
and many composite populations thus provide adequate features.
Once the gap in the data between 0.95 and 1.95\,$\mu$m will have
been filled, the shape of the complete near-IR spectrum, that
is determined to a large extent by molecular features, should
rule out some of the models we must still take as acceptable here.

It is instructive to locate the global and central data for NGC\,7714
on the diagram of Meurer et al. (1999), which shows a correlation
between the ultraviolet spectral index $\beta$ ($f_{\lambda} \propto
\lambda^{\beta}$ in the IUE spectral range, as defined by Calzetti et al.
1994) and the ratio of far-IR to UV flux, for large aperture observations
of a sample of starburst galaxies of various types. This is done in 
Fig.\,\ref{Meurer_plot.fig}. When observed through large apertures,
NGC\,7714 follows the main trend (Meurer et al., 1999). 
Recall that about half the IUE flux comes
from the central $\sim$330\,pc, and that 10 to 50\,\% of the 
far-IR flux originates there. This places the central region
off the relation, further than any of the global galaxy data of the sample
of Meurer et al. (1999). For NGC\,7714, the global properties
clearly are the result of large scale averaging of very diverse local 
properties. Along individual lines of sights, random effects
of the dust distribution are large; detailed UV to far-IR
energy budgets are made uncertain by unknown losses and gains through 
scattering on gas and dust. We extrapolate this conclusion, 
and suggest that the smoothing effect of observations through
large apertures plays an important role in producing the relation
of Meurer et al. (1999). 

Should one expect searches to find whole galaxies at locations in the
diagram close to NGC\,7714's central regions? 
Obtaining as negative a value of $\beta$ would require a very
prominent young population to lie essentially in front of any
dust or in a dust hole; this is not impossible, but probably rare.
Objects above the main trend at $\beta \simeq -1$ would be less surprising,
as that is an expected location when significant amounts
of dust are mixed among the stellar populations: at each wavelength, one 
then sees down to optical depths of order one, which results in
$\beta \simeq -1$, rather independently of the (large) far-IR 
emission. Meurer (2000) indeed finds that the ultraluminous infrared starburst
galaxies lie in that area, with the objects studied to date having
$-2\leq \beta \leq -0.5$ and $\log(F_{\mathrm FIR}/F_{\mathrm FUV}) \geq 2$; 
we would not be surprised if the gap between the two current galaxy samples 
in the diagram were eventually filled.

\section{Conclusion}
\label{Concl.sec}

A supposedly simple system as the proto-typical starburst nucleus
in NGC\,7714 turns out to be rather complex. High resolution imaging 
reveals highly irregular structure even within the central $\sim 330$\,pc.
In the study of the energy distribution of this region, star formation
history, geometry and grain properties conspire against evolutionary
synthesis models. The degeneracy is only partly broken even when data
throughout the electromagnetic spectrum are combined. However, several
significant results follow from the collection of models that 
satisfactorily reproduce the photometric and spectroscopic data:

\begin{itemize}

\item The nuclear $\sim 330$\,pc of NGC\,7714 have been forming stars off and
on over the past several hundred Myr, at an average star formation rate
of the order of 1\,M$_{\odot}$\,yr$^{-1}$. About 5\,Myr ago, an enhancement
over this average rate by a factor of a few has occurred.

\item The mass of gas transformed into stars over the last several 100\,Myr
amounts to $1-5 \times 10^8$\,M$_{\odot}$ (in the $1-80$\,M$_{\odot}$
mass range, with a Salpeter IMF). Of these, 
$4-20 \times 10^6$\,M$_{\odot}$ were processed within the most recent star 
formation episode (the exact value depends sensitively on the dust 
distribution).

\item The duration of the extended star formation episode determined 
spectrophotometrically is longer than the 110\,Myr elapsed since 
closest contact with NGC\,7715, according to the dynamical study
of Smith \& Wallin (1992). Only a small subsample of our 
satisfactory models is consistent with a timescale of 200\,Myr,
that would be within the uncertainties of the dynamical estimate.
While the last interaction may have triggered the most recent
star formation, it is unclear what first initiated starburst activity.

\item The spectrophotometric data place (model dependent) limits on the
contribution of the underlying population. Assuming the 
same IMF as for the burst and a standard spiral star formation law, we
find an upper limit of about 10$^9$\,M$_{\odot}$ for this contribution.
Thus, the recent star formation episodes have enhanced the
stellar mass in the nucleus by at least 10\,\% (most models suggesting
about 25\,\%).

\item The UV luminosity is due to leakage of a small fraction
 ($\sim$10\%) of the stellar photons, from a central cluster that happens
 to be unobscured, or from a central hole in the ISM.

\item In the GHRS aperture, the stars younger than $\sim$5\,Myr (among which
 those actually seen in the UV), do not suffice to explain the far-IR emission.
 A large contribution from intermediate age stars
 ($\sim 10^7$\,yrs old and not seen in the UV) or, alternatively,
 significant photon exchanges with regions outside the
 aperture (through scattering on gas or dust) are essential.

\end{itemize}

The HST images of the center of NGC\,7714 show $10-20$ individual bright
spots, possibly clusters. None of these is dominant (in bolometric
light) to be identifiable as the galaxy nucleus. The stellar populations
that dominate at UV, optical and near-IR wavelengths have different ages
and different amounts of extinction. This spectrophotometric result
is in excellent agreement with multiwavelength high resolution imaging,
which shows, for instance, that the UV continuum and the Br$\gamma$
line emission maxima do not coincide in space. It is not surprising
that values of $E(B-V)$ determined from 
Balmer lines or from the continuum at various wavelengths
differ. The light contribution of the underlying population is
strongest in the near-IR, where the patchy effects of extinction
are also lowest. These facts together explain the relatively smooth
appearance of the nucleus in the near-IR continuum.

NGC\,7714 has been included in almost all samples of 
nuclear starburst galaxies, and is considered the archetypal UV-bright 
case. Taken as a whole, NGC\,7714 lies on the relation between
far-IR luminosity and UV continuum slope found by  Meurer et al. (1999)
for a sample of large aperture observations of UV-detected starbursts.
Our analysis has shown that the central $\sim 330$\,pc lie off
this correlation, because of a few lines of sight with little 
obscuration towards hot stars. Most of the stars in this small aperture
and in larger ones are more severely obscured. The global
properties of NGC\,7714, as probably those of many starburst galaxies, result 
from the averaging over many lines of sight with very diverse
individual obscuration properties.

Incidentally, both M~82 and NGC~7714 have similar
bolometric luminosities, indicating similar star-formation rates. What
distinguishes the two galaxies is the dust column along the line of
sight. M~82 is viewed almost edge-on, leading to a large dust column,
and therefore large obscuration, whereas NGC~7714 is seen closer to
pole-on, and the partially naked core is exposed. We can speculate that
NGC~7714 would resemble M~82 if its inclination were similar.

What lessons have we learned from NGC\,7714 for the derivation of 
star formation rates from the UV or the far-IR light? In agreement with
Paper~I, the analysis of the nuclear UV data alone indicates about 
$5\times 10^6$\,M$_{\odot}$ of $\sim 5$\,Myr young stars. The total
mass of stars of this most recent episode of star formation is most likely
higher, by a factor of 2 to 4 (the Br$\gamma$ emission
carries the strongest evidence for the presence of obscured O,B stars,
and the whole spectrum was necessary to set constraints on their
number). The blue UV continuum of the nucleus of NGC\,7714 is
reproduced by the models as long as about 25\,\% of the hot stars is 
obscuration free.  On a larger spatial scale, or
in a more massive galaxy with a higher dust content, it becomes
less likely that as much as 25\,\% of the young stars may happen to
be obscuration free. The UV continuum will appear reddened, as it
is the case for NGC\,7714 in the IUE aperture and for most galaxies
of the sample of Meurer et al. (1999). However, as most clearly
demonstrated by Witt \& Gordon (2000), the change in UV slope only
gives a lower limit to the actual attenuation of the UV light. IRAS
galaxies with relatively blue UV continua confirm this bias
(Meurer 2000). 

While the UV light in NGC\,7714  traces
the most recent star formation to within a factor of a few, it 
clearly is blind to the $\sim 10^8$\,M$_{\odot}$ of stars that were 
born over the larger timespan of a few $10^8$\,yrs. Our models show
that the latter (in particular the stars born $10^7-10^8$\,yrs
ago) contribute a large fraction of the far-IR emission of the 
area. Thus, estimates from UV or far-IR
measurements provide star formation rates averaged over different
timescales. Information about the total duration of the star
formation episode is necessary to estimate the real importance
of a starburst.  Gordon et al. (2000) show that the far-IR to UV
flux ratio is a safe indicator of the UV attenuation for simple
stellar populations. But for composite populations such as those
of real starbursts we confirm the author's warnings that even this method 
is hazardous.

\acknowledgments
\noindent Acknowledgments:\\
J.G. performed most of his work on this paper while holding a postdoctoral
fellowship at Space Telescope Science Institute. The STScI
is also thanked for its hospitality to A.L. for two visits that
made efficient progress possible.
C.L. wishes to thank Observatoire de Strasbourg (Universit\'e
Louis Pasteur) for its kind hospitality and support 
when this paper was finalized. This work was supported by
HST grant GO-06672.01-95A from the Space Telescope Science Institute,
which is operated by the Association of Universities for Research in
Astronomy, Inc., under NASA contract NAS5-26555.




\clearpage



\figcaption[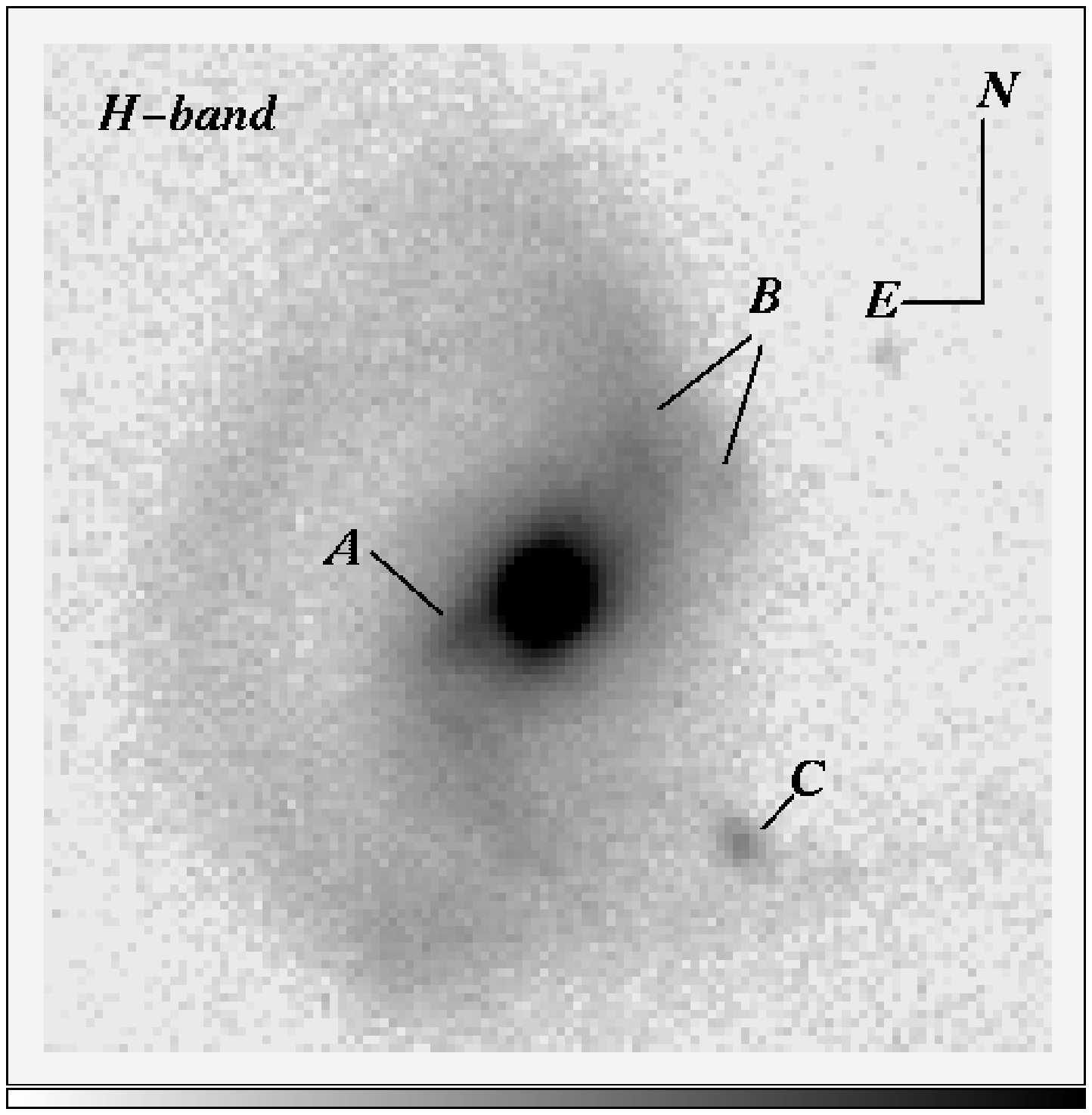]{\label{caspir} H-band image of NGC~7714 taken
with the CASPIR camera.  The field size is 1$'$. Regions are labelled 
as in Gonz\'alez-Delgado et al. (1995).}

\figcaption[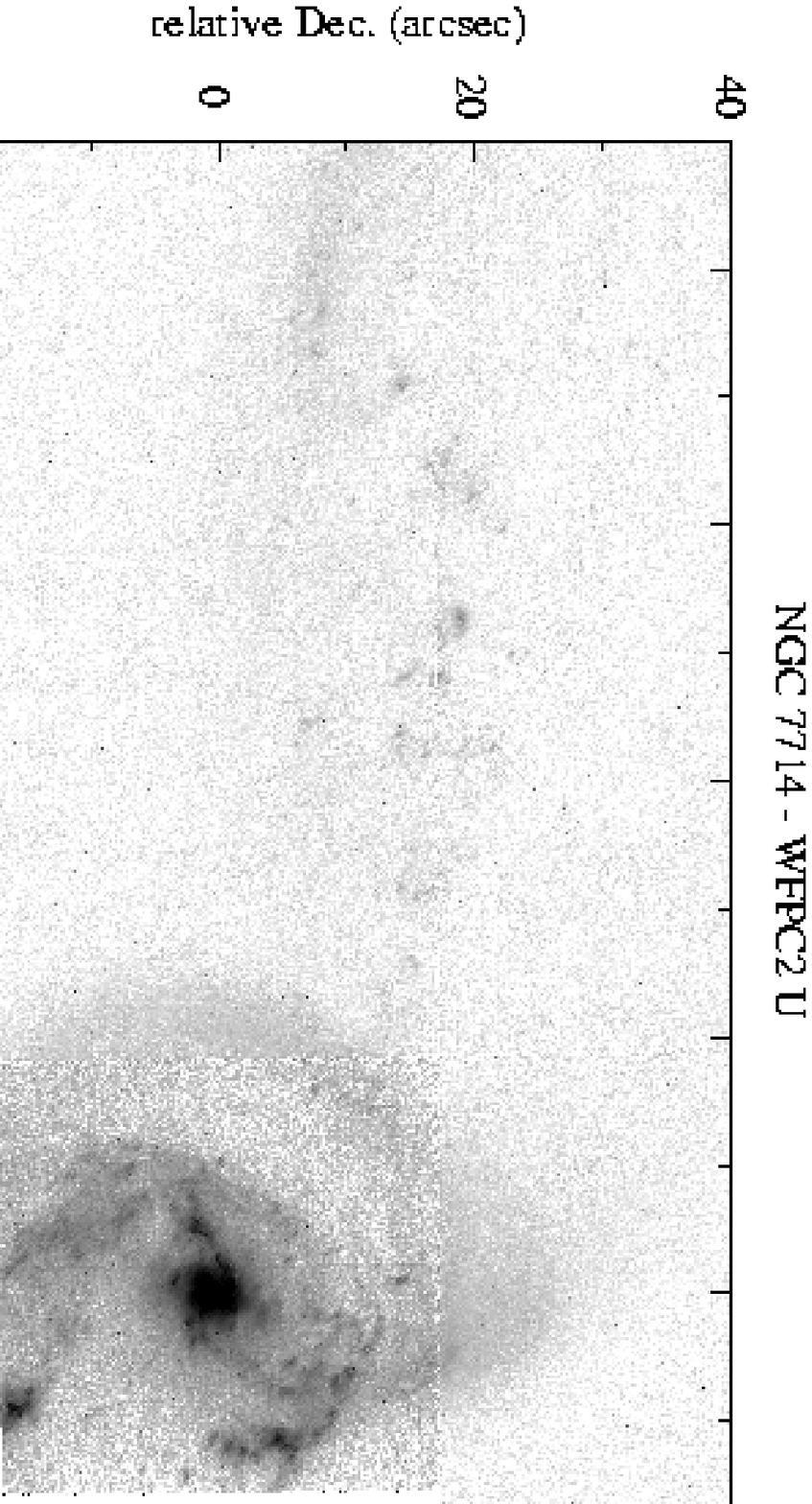]{\label{uvwfpc2a} Mosaic of the whole galaxy in 
the F380W filter, through WFPC2.  The scale is 0.1$\arcsec$/pixel; North is
up, East to the left.  The bridge
between NGC~7714 and its companion (barely detected) is visible.}

\figcaption[f3.eps]{\label{uvwfpc2b} Nuclear zone of NGC~7714
in the PC1 chip of WFPC2, in the F380W (upper left) and F606W (upper 
right) filters.  The scale is  0.046$\arcsec$ per pixel.  North is up and East 
is to the left.  We also show the ratio F606W/F380W (lower left), 
indicating that most of the clusters are bluer 
than their surroundings (light=blue, dark=red).  In the lower right
panel, we show an unsharp-masked
image in F380W, which emphasizes the dark lanes in the nucleus.  Note
that the dark lanes are somewhat red in the color image, and that
one of the lanes passes over the apparent geometric center of the nucleus.}

\figcaption[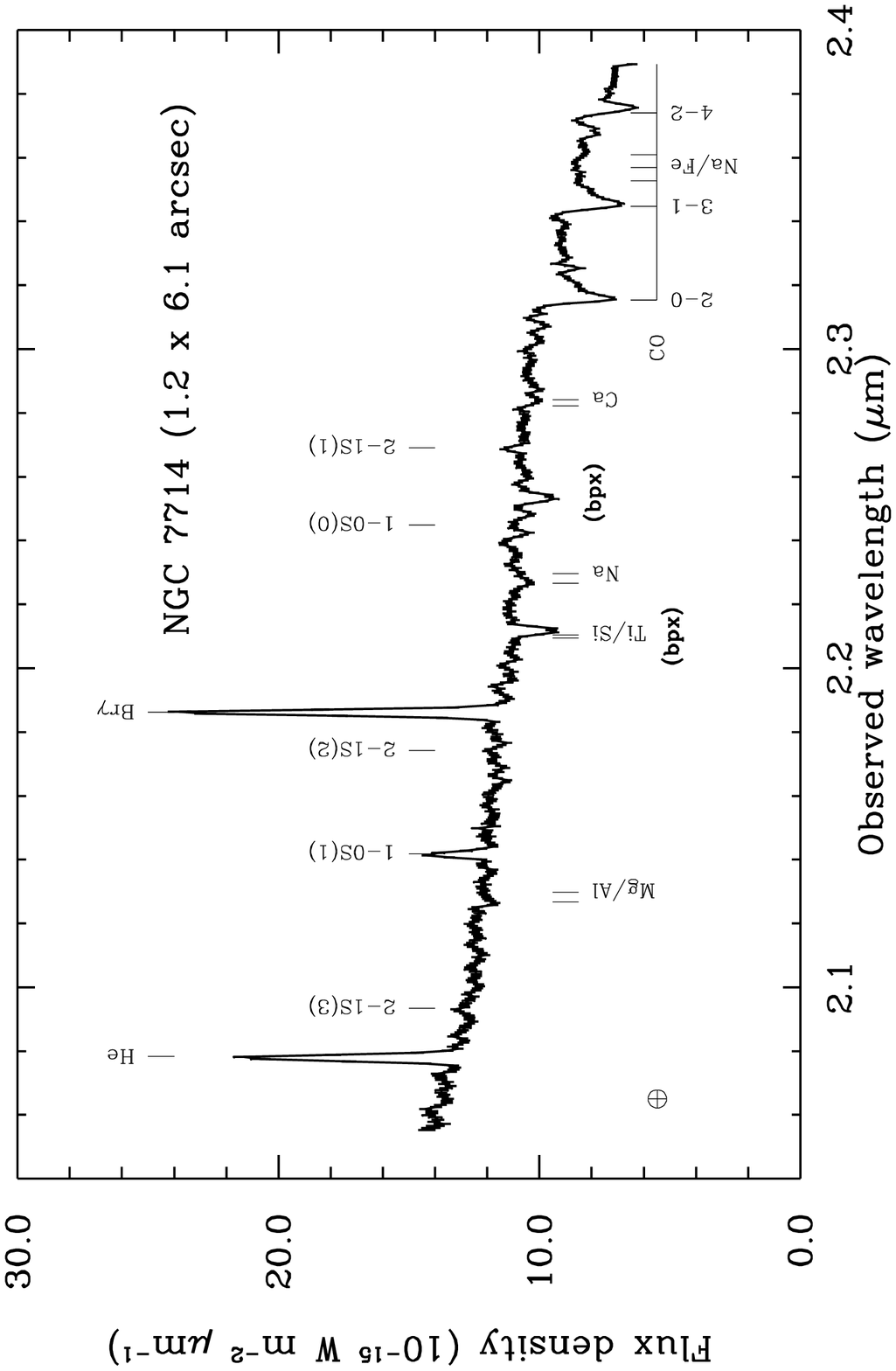]{\label{centspec} IR spectrum 
in an aperture of 1~$\times$~5 pixels.}

\figcaption[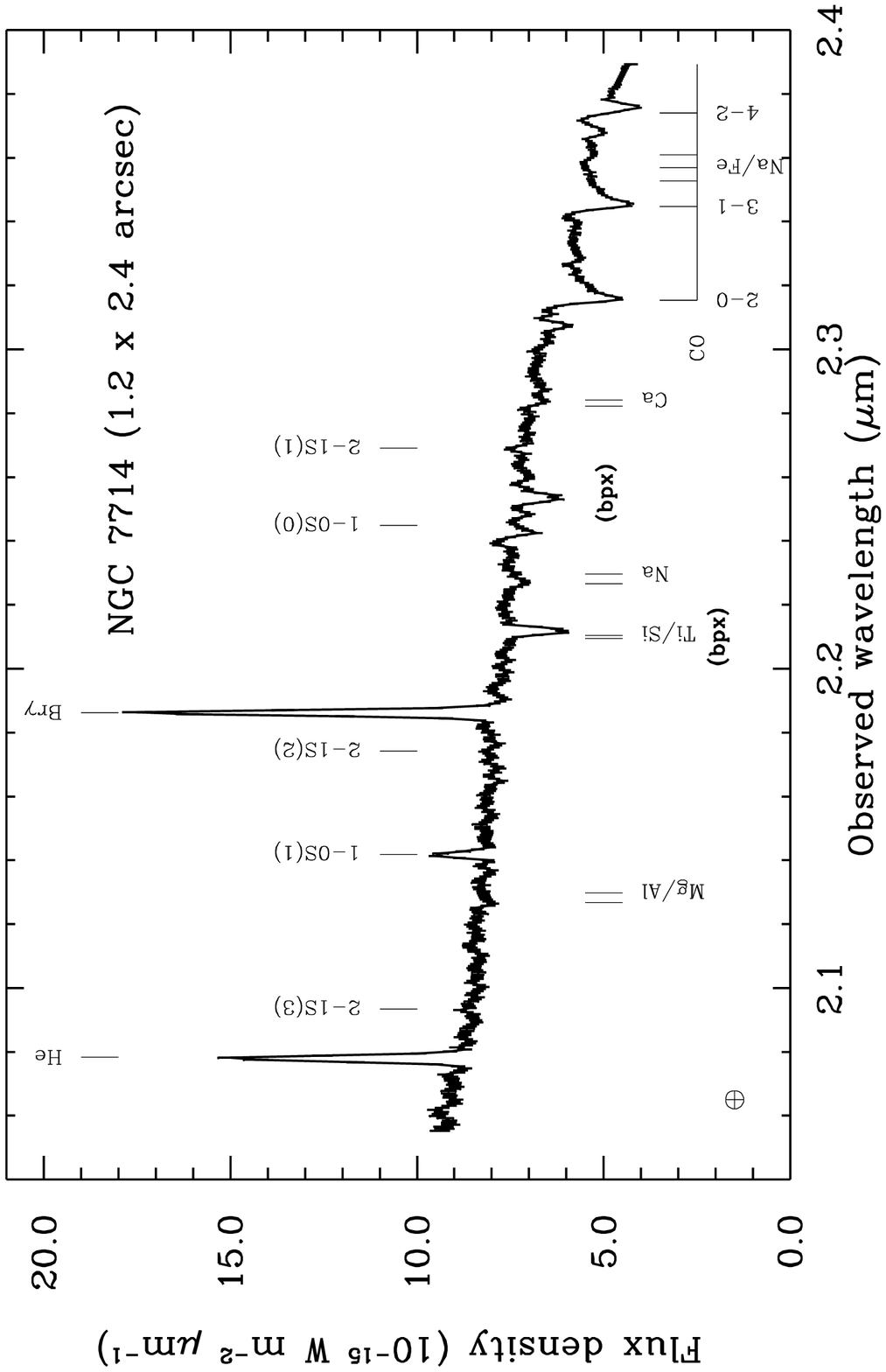]{\label{nucspec} IR spectrum 
of the nucleus itself ($1.2\arcsec \times 2.4\arcsec$ aperture).}

\figcaption[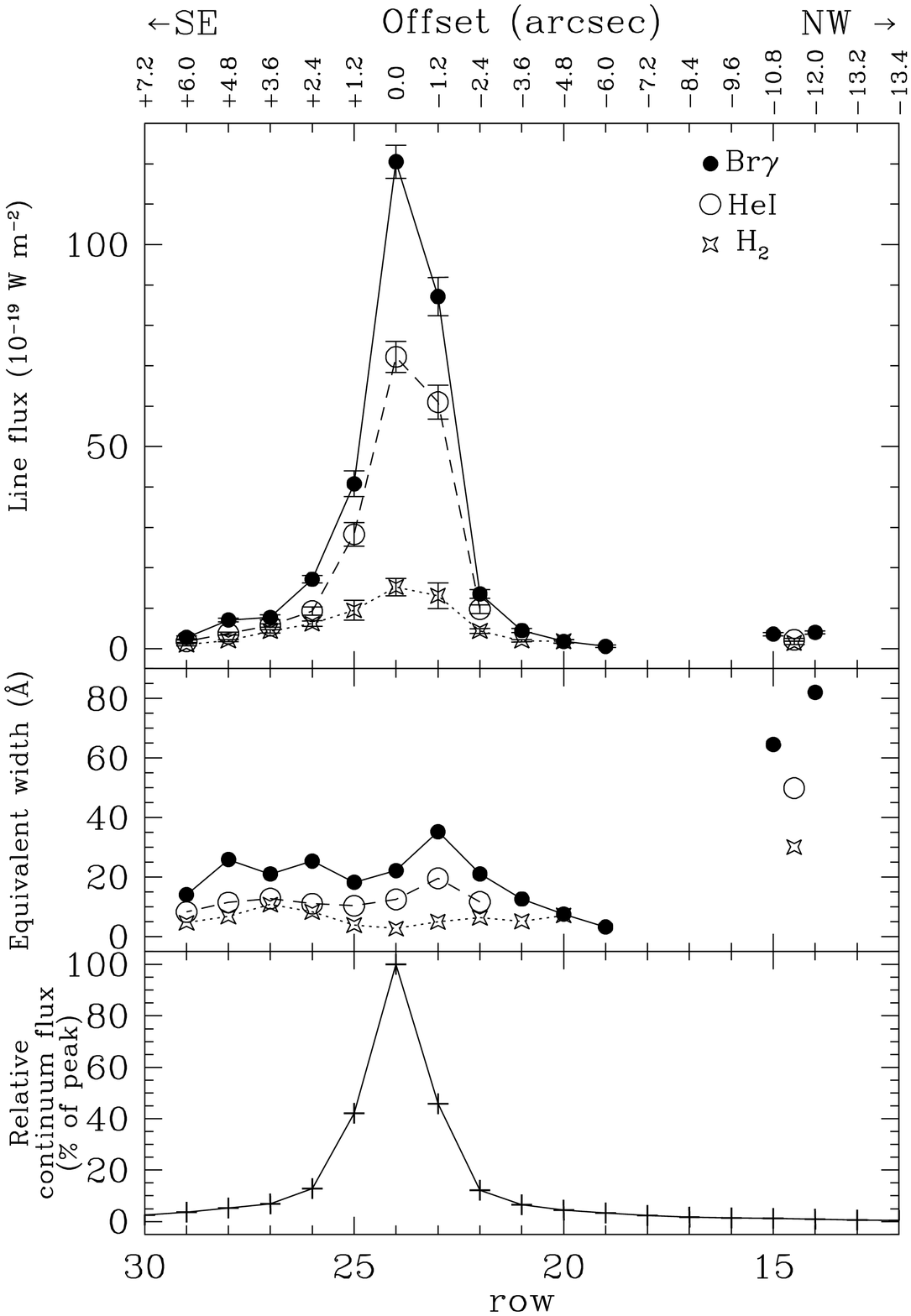]{\label{lines} Line fluxes and equivalent 
widths as a function of position along the slit.  The top panel shows the 
fluxes of the Br$\gamma$, He~I, and H$_2$1--0 S(1) lines in individual 
spatial rows.  The middle panel shows the equivalent widths
in individual rows.  In the top and middle panels,
the rightmost points for the He and H$_2$ lines are the sums of the
lines in rows 14 and 15 of the spectrum.  The bottom panel shows the 
spatial profile of the continuum emission.  The slit passes across the
nucleus (peaking at row 24),
H~{\sc ii} region B in rows 14 and 15, and H~{\sc ii} region A at approximately
row 28.}

\figcaption[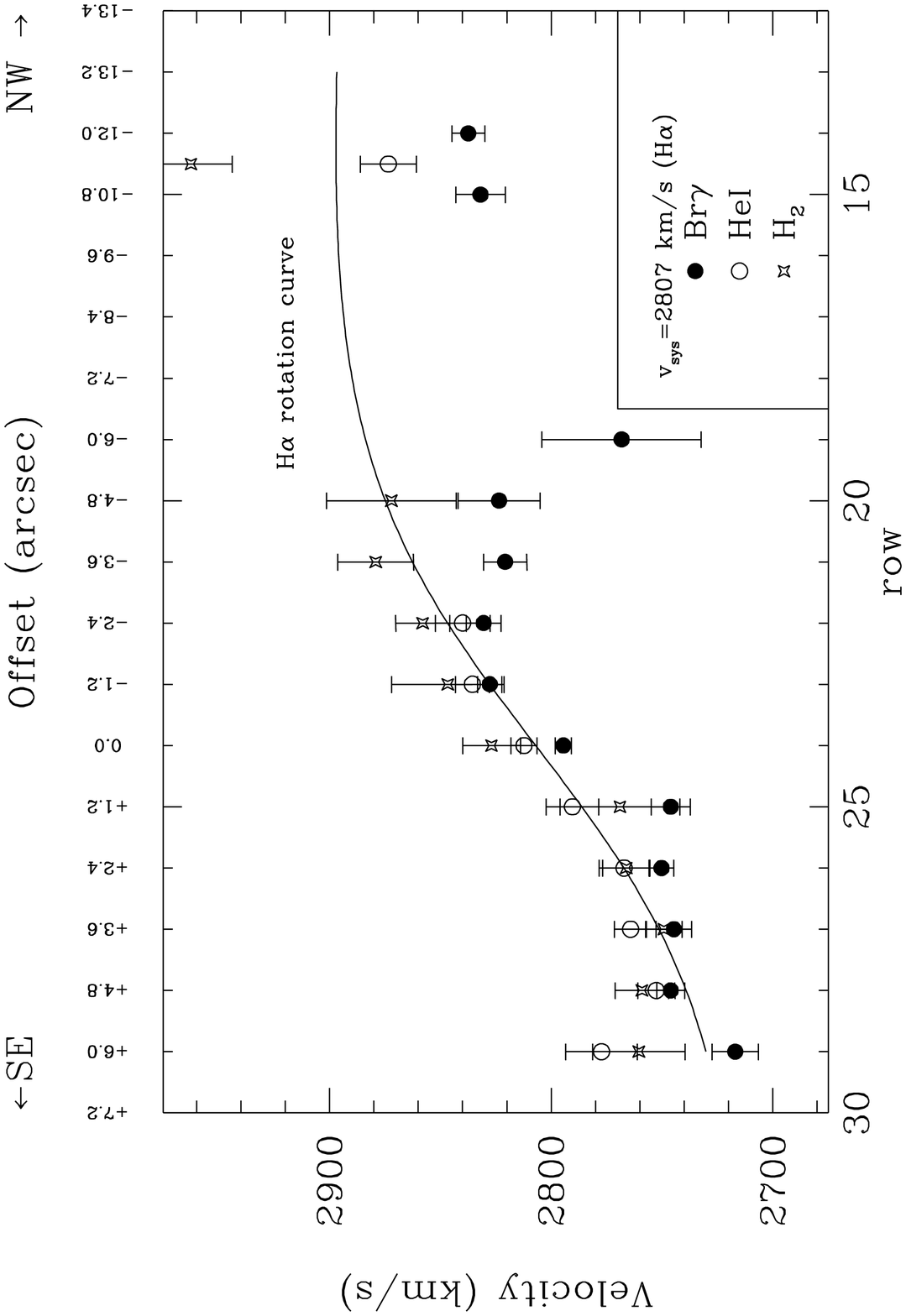]{\label{kinematics} Observed velocities
 of the IR emission lines as function of position along the slit, along 
with the rotation curve
derived for the H$\alpha$ line.  The rotation curve is given
after \citet{gon95} and \citet{ber91} as:
as $v(r)=\frac{Ar}{(r^2+C_0^2)^{p/2}}$, where $A=231$~\kms, $C_0=8.3\arcsec$,
and $p=1.35$, for a spherical potential model with pure circular motions.  
As in \citet{gon95}, corrections have been made for both the inclination 
(assumed to be 45$^\circ$), and the difference between the slit position angle 
and the major
axis (assumed to be 35$^\circ$).}

\figcaption[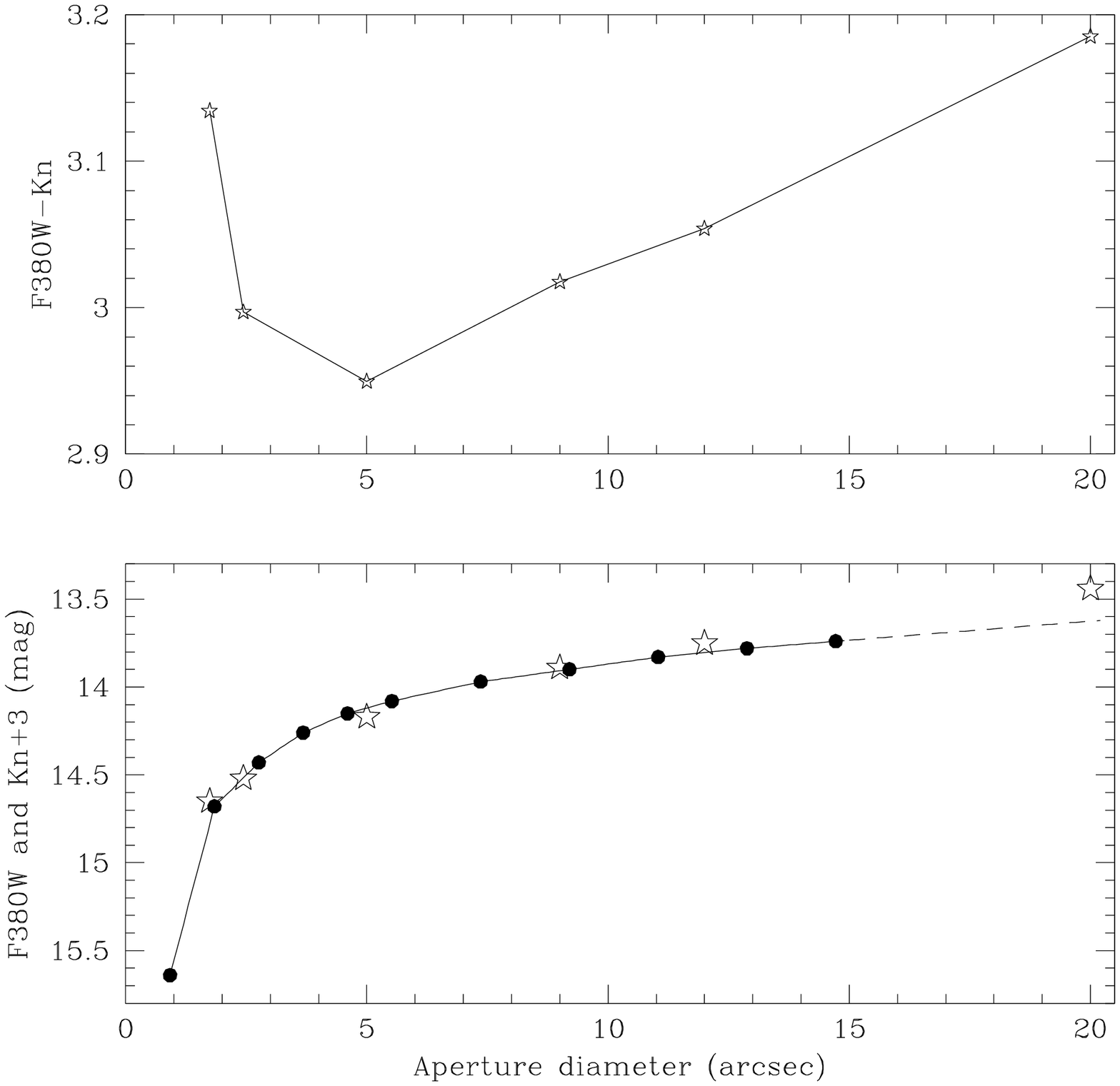]{\label{aperture_col_lin}
Evolution of the UV and near-IR integrated magnitudes and of 
the corresponding color with aperture size.}

\figcaption[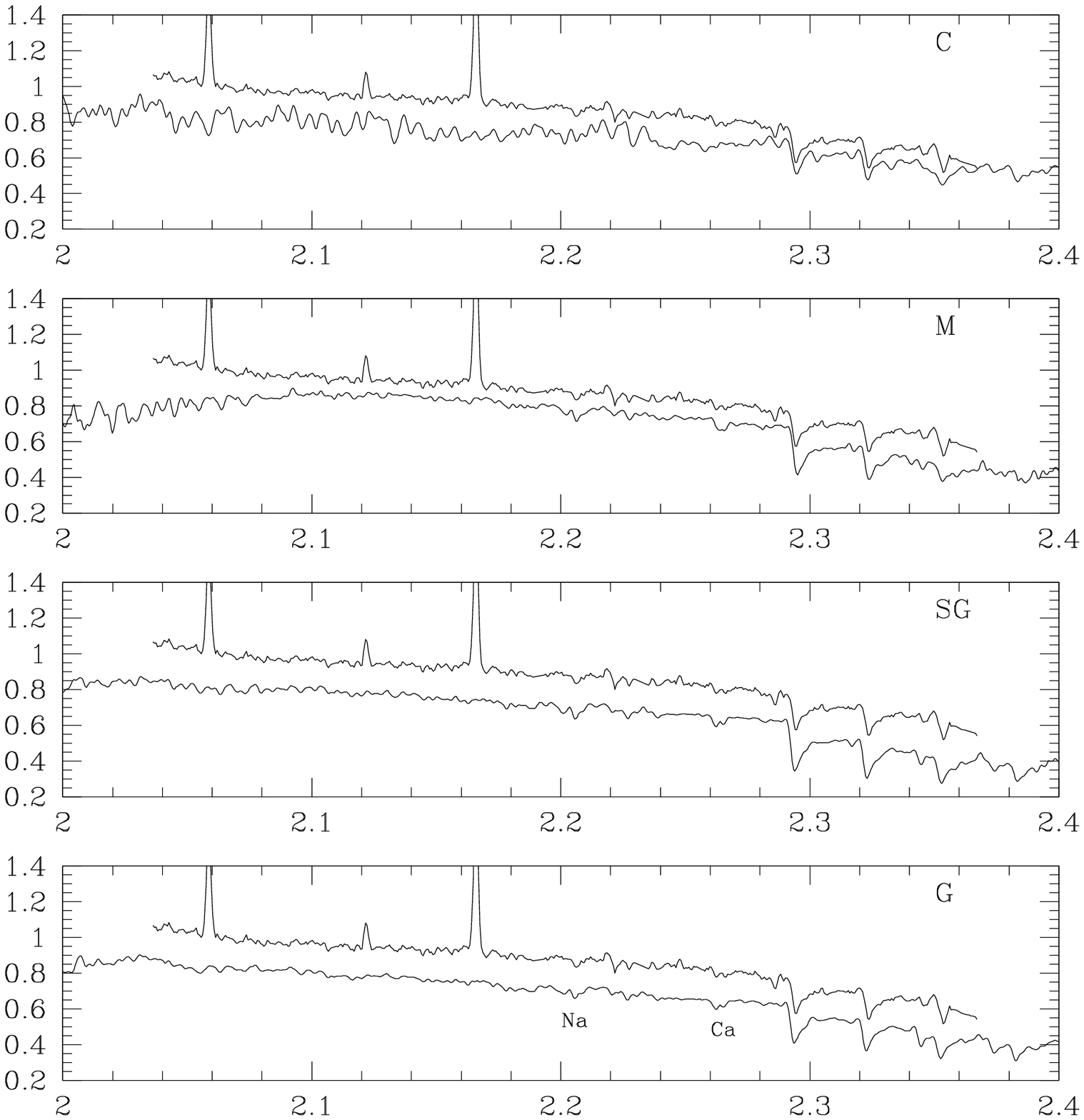]{\label{N7714_IR_stars.fig}
The near-IR spectrum of NGC\,7714, compared, from bottom to top,
with the spectrum of a red giant, a red supergiant, an oxygen-rich
variable AGB star and a carbon-rich AGB star. The spectra have 
been normalized to the same flux at 2.2\,$\mu$m, then shifted
for display purposes.}

\figcaption[f10.eps]{\label{1pop.2screens.fig}
Illustrative fit of the nuclear energy distribution of NGC\,7714 with
a simple, homogeneous star formation history, but an
inhomogeneous dust distribution. Here, star formation has 
occured constantly for 1\,Gyr; $E(B-V)=0$ for some lines of
sight, 0.45 for others. The models shown are computed with
P\'egase and have the spectral resolution of the stellar library of
Lejeune et al. (1997).}

\figcaption[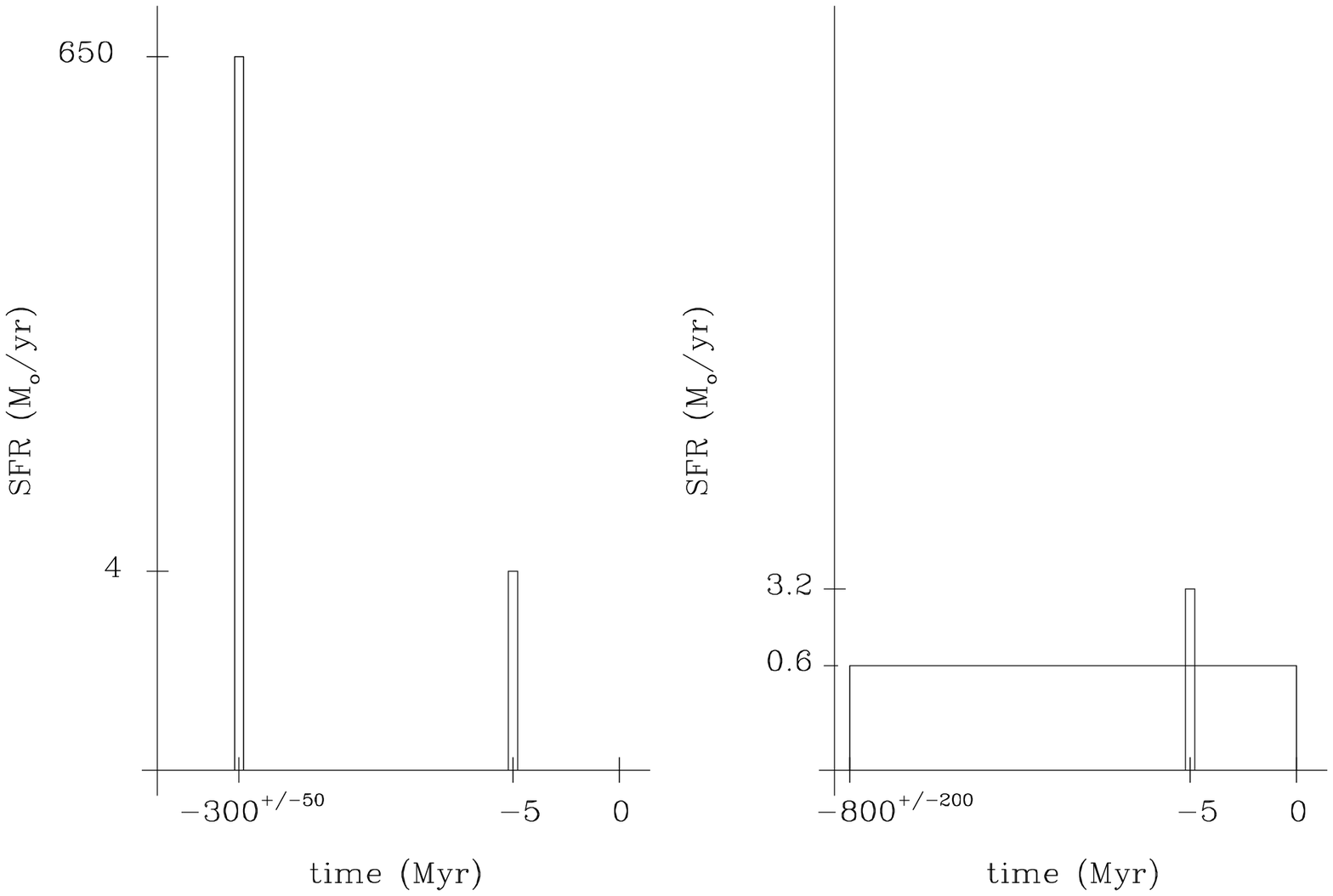]{\label{2pop.schematic.fig}
Schematic representation of the recent star formation rates for 
two-component models 
that reproduce the UV-V-NIR nuclear spectrum of NGC\,7714 rather well (right)
and marginally well (left). However, these models fail to satisfy constraints
outside this wavelength range (see text).}

\figcaption[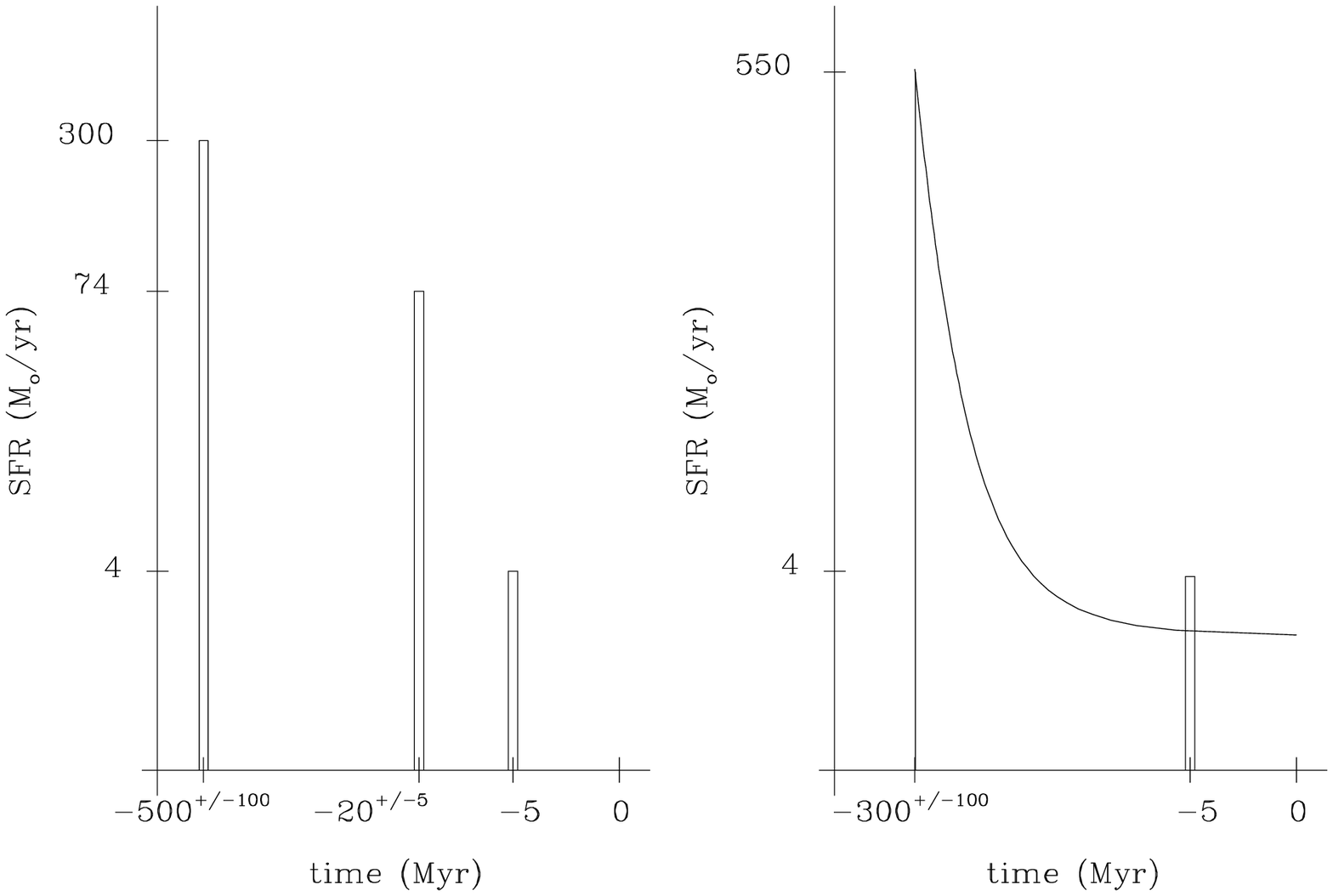]{\label{3pop.schematic.fig}
Two possible scenarios for the star formation in the nucleus of NGC\,7714 over
the last few 100 Myr.  The star formation rate
has been globally decreasing. The 5 Myr burst currently seen in the UV is 
part of a larger star formation scheme.}

\figcaption[f13.eps]{\label{3pop.fit.fig} 
Illustrative fit of the nuclear spectrum with 3 successive bursts of decreasing
burst masses. Dashed: burst age\,=\,5 Myr, intrinsic $E(B-V)=0.03$, 
$M=4 \times 10^6$~\Ms\ (with the assumed 1~--~80\,\Ms\ Salpeter IMF). 
Dotted: burst age\,=\,20 Myr, intrinsic $E(B-V)=0.3$,
$M=7 \times 10^7$~\Ms. Dot-dashed: burst age\,=\,500\,Myr,
intrinsic $E(B-V)=0.15$, $M=3 \times 10^8$~\Ms. 
$3.5 \times 10^8$~\Ls, $4 \times 10^9$~\Ls, and $4 \times 10^8$~\Ls\ 
of the light emitted by
these components
are absorbed by dust, resulting in a total contribution of
 $4.8 \times 10^9$~\Ls\
to the IRAS emission of the galaxy.}

\figcaption[f14.eps]{\label{3pop.underl.fig}
Illustrative fit of the nuclear spectrum, maximizing
 the contribution of the underlying spiral galaxy 
population to the near-IR light and minimizing the age of the recent
star formation episode. Dashed: 
as in Fig.\,\ref{3pop.fit.fig}, with a stellar mass of nearly
$4\times10^6$~\Ms. Dotted: spiral galaxy 
population whose star formation rate followed a Schmidt law for 10\,Gyr
(SFR~=~0.3~\Myr\ $\times$  available gas mass, closed box), thus 
having produced about $7 \times 10^8$~\Ms\ of stars (in the 1--80\,\Ms\ 
mass range); 
intrinsic $E(B-V)=0.29$.
Dot-dashed: intermediate age population (exponentially decreasing SFR
with an $e$-folding time of 50\,Myr and an age of 200 Myr, $E(B-V)=0.2$),
$M=1 \times 10^8$~\Ms.
$3.3 \times 10^8$~\Ls,   $1.3\times 10^9$~\Ls,
and  $1.6 \times 10^9$~\Ls\
of the light emitted by these components
are absorbed by dust, providing a total contribution of $3.2\times 10^9$~\Ls\
to the IRAS emission of the galaxy. }

\figcaption[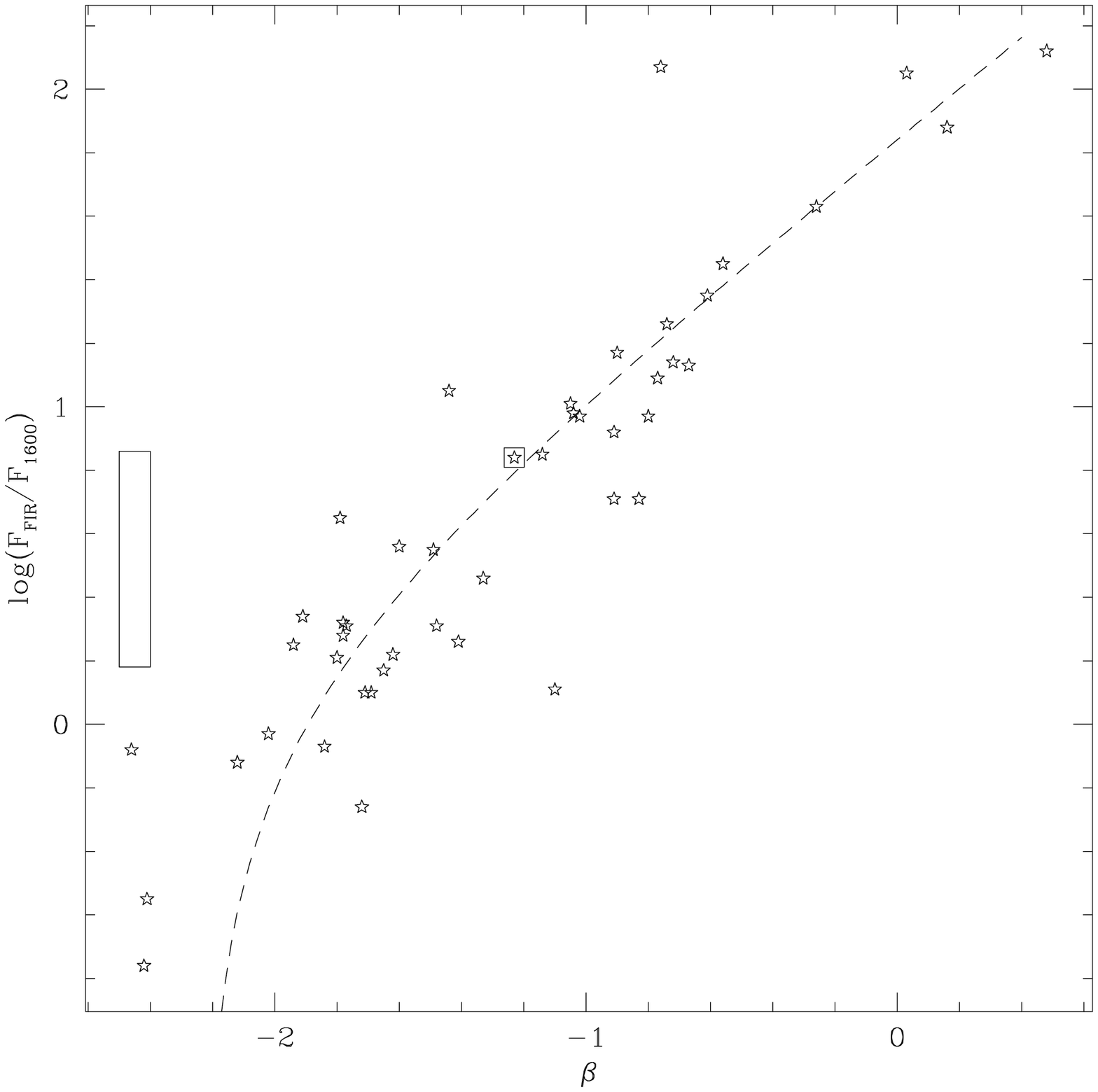]{\label{Meurer_plot.fig}
Correlation between global far-IR to UV flux ratios and the UV
spectral index $\beta$. The stars and the dashed line are the data and
the suggested relation of Meurer et al. (1999; their Fig.\,1)
for their sample of galaxies of various types. The square identifies 
NGC\,7714 as observed through a large aperture. 
The rectangle shows the location of the nuclear region
of NGC\,7714 studied in this paper. }

\clearpage

\psfig{figure=f1.eps,angle=0,width=17cm,height=17cm}
Fig.~\ref{caspir}
\clearpage

\psfig{figure=f2.eps,angle=180,width=14cm,height=20cm}
Fig.~\ref{uvwfpc2a}
\clearpage

\psfig{figure=f3.eps,angle=-90,width=17cm,height=17cm}
Fig.~\ref{uvwfpc2b}
\clearpage

\psfig{figure=f4.eps,angle=-90,width=17cm,height=15cm}
Fig.~\ref{centspec}
\clearpage

\psfig{figure=f5.eps,angle=-90,width=17cm,height=15cm}
Fig.~\ref{nucspec}
\clearpage

\psfig{figure=f6.eps,angle=0,width=15cm,height=20cm}
Fig.~\ref{lines}
\clearpage

\psfig{figure=f7.eps,angle=-90,width=17cm,height=15cm}
Fig.~\ref{kinematics}
\clearpage

\psfig{figure=f8.eps,angle=0,width=16cm}
Fig.~\ref{aperture_col_lin}
\clearpage

\psfig{figure=f9.eps,angle=0,width=15cm,height=18cm}
Fig.~\ref{N7714_IR_stars.fig}
\clearpage

\psfig{figure=f10.eps,angle=-90,width=17cm,height=15cm}
Fig.~\ref{1pop.2screens.fig}
\clearpage

\psfig{figure=f11.eps,angle=0,width=15cm,height=18cm}
Fig.~\ref{2pop.schematic.fig}
\clearpage

\psfig{figure=f12.eps,angle=0,width=15cm,height=18cm}
Fig.~\ref{3pop.schematic.fig}
\clearpage

\psfig{figure=f13.eps,angle=-90,width=17cm,height=15cm}
Fig.~\ref{3pop.fit.fig}
\clearpage 

\psfig{figure=f14.eps,angle=-90,width=17cm,height=15cm}
Fig.~\ref{3pop.underl.fig}

\psfig{figure=f15.eps,angle=0,width=15cm}
Fig.~\ref{Meurer_plot.fig}


\clearpage

\begin{table}
\begin{center}
\caption[]{Near-IR Aperture Photometry of NGC 7714}
\label{NIRphot.tab}

\myvspace

\begin{tabular}{cccccccc} \hline \hline
~  &  
\multicolumn{3}{c}{This study} &
\multicolumn{3}{c}{Literature} &
Reference  \\
Aperture  & $J$ & $H$ & $Kn$ & $J$ & $H$ & $K$ & \\
diameter  & (mag) & (mag) & (mag) & (mag) & (mag) & (mag) & \\
\hline

1.74$\arcsec$ & 12.66 & 11.99 & 11.65 & ... & ... & ... & \\
2.44$\arcsec$ & 12.51 & 11.85 & 11.52 & ... & ... & ... & \\
5$\arcsec$    & 12.13 & 11.47 & 11.17 & 12.27 & 11.56 & 11.23 & $a$ \\
9$\arcsec$    & 11.82 & 11.12 & 10.89 & 11.97 & 11.21 & 10.89 & $b$ \\
12$\arcsec$   & 11.72 & 11.02 & 10.75 & 11.81 & 11.08 & 10.77 & $b$ \\
20$\arcsec$   & 11.39 & 10.71 & 10.44 & ... & ... & ... & \\
34$\arcsec$   & 11.18 & 10.49 & 10.25 & 11.16 & 10.41 & 10.10 & $b$ \\
\hline
\end{tabular}
\end{center}

\myvspace

$a$- Lawrence et al. (1985);
$b$- Glass \& Moorwood (1985).
\end{table}

\clearpage

\begin{table}
\begin{center}
\caption[]{F380W Aperture Photometry of NGC 7714}
\label{Uphot.tab}

\myvspace

\begin{tabular}{cc} \hline \hline
Radius & Magnitude \\ \hline
0.46$\arcsec$ & 15.64 \\
0.92$\arcsec$ & 14.68 \\
1.38$\arcsec$ & 14.43 \\
1.84$\arcsec$ & 14.26 \\
2.30$\arcsec$ & 14.15 \\
2.76$\arcsec$ & 14.08 \\
3.68$\arcsec$ & 13.97 \\
4.60$\arcsec$ & 13.90 \\
5.52$\arcsec$ & 13.83 \\
6.44$\arcsec$ & 13.78 \\
7.36$\arcsec$ & 13.74 \\ \hline
\end{tabular}
\end{center}
\end{table}

\clearpage

\begin{table}
\begin{center}
\caption[]{Selected K-band Line Fluxes in NGC 7714}
\label{NIRlines.tab}

\myvspace

\begin{tabular}{lc} \hline \hline
\multicolumn{1}{c}{Line} & Flux$^a$ \\
     & (10$^{-19}$\,W\,m$^{-2}$) \\
\multicolumn{1}{c}{(1)} & (2) \\ 
\hline

\multicolumn{2}{c}{Aperture $1.22\arcsec\,\times\,6.10\arcsec$ (Central)} \\ 
\hline

He{\sc i} 	& 191.0\,$\pm$\,6.8	\\
H$_2$\,2--1\,S(3)	& $<$ 17.2  (3$\sigma$) \\
H$_2$\,1--0\,S(1)	& 40.9\,$\pm$\,4.3 \\
H$_2$\,2--1\,S(2)	& $<$ 16.0  (3$\sigma$)  \\
Br$\gamma$	&  274.5\,$\pm$\,7.4  \\
H$_2$\,1--0\,S(0)	& $<$ 13.8  (3$\sigma$)  \\
H$_2$\,2--1\,S(1)	& 18.1\,$\pm$4.0  \\
\hline

\multicolumn{2}{c}{Aperture $1.22\arcsec\,\times\,2.44\arcsec$ (Nuclear)} \\ \hline

He{\sc i}       	&  139.0\,$\pm$\,5.8     \\
H$_2$\,2--1\,S(3)       & $<$ 14.4  (3$\sigma$) \\
H$_2$\,1--0\,S(1)       & 27.2\,$\pm$\,3.7 \\
H$_2$\,2--1\,S(2)       & $<$ 13.6  (3$\sigma$)  \\
Br$\gamma$      	&  209.6\,$\pm$6.2  \\
H$_2$\,1--0\,S(0)       & $<$ 11.4  (3$\sigma$)  \\
H$_2$\,2--1\,S(1)       & 8.4\,$\pm$3.1  \\
\hline

\multicolumn{2}{c}{Two outer pixels} \\ \hline

He{\sc i}               &  18.6\,$\pm$\,1.5     \\
H$_2$\,2--1\,S(3)       & $<$ 4.3  (3$\sigma$) \\
H$_2$\,1--0\,S(1)       & 10.6\,$\pm$\,1.0 \\
H$_2$\,2--1\,S(2)       & $<$ 3.4  (3$\sigma$)  \\
Br$\gamma$              &  29.4\,$\pm$1.4  \\
H$_2$\,1--0\,S(0)       & 5.3\,$\pm$0.9  \\
H$_2$\,2--1\,S(1)       & 3.1\,$\pm$0.9  \\
\hline

\end{tabular}
\end{center}

\myvspace

$a$. The quoted errors are from the fitting procedure and do not 
include the estimated 10\% calibration uncertainty.

\end{table}

\clearpage

\begin{table}
\begin{center}
\caption[]{H$_2$ Line Ratios in NGC 7714}
\label{H2lines.tab}

\myvspace

\begin{tabular}{lcccc} \hline \hline

\multicolumn{1}{c}{Aperture} &
\multicolumn{4}{c}{Ratio} \\
~ & $\frac{2-1\,\rm{S}(3)}{1-0\,\rm{S}(1)}$ &
$\frac{2-1\,\rm{S}(2)}{1-0\,\rm{S}(1)}$ &
$\frac{1-0\,\rm{S}(0)}{1-0\,\rm{S}(1)}$ &
$\frac{2-1\,\rm{S}(0)}{1-0\,\rm{S}(1)}$ \\
\multicolumn{1}{c}{(1)} & (2) & (3) & (4) & (5) \\ \hline

``Central" spectrum ($1.2\arcsec\times 6.1\arcsec$) 	&
	$<$ 0.42 & $<$ 0.39 & $<$ 0.34 & 0.44\,$\pm$\,0.11  \\
``Nuclear" spectrum ($1.2\arcsec\times 2.4\arcsec$)	&
	$<$ 0.53 & $<$ 0.50 & $<$ 0.42 & 0.31\,$\pm$\,0.12  \\
Two outer pixels 	&
	$<$ 0.41 & $<$ 0.32 & 0.50\,$\pm$\,0.10 & 0.29\,$\pm$\,0.09   \\
Black \& van Dishoeck model 14   	&
	0.35 & 0.28 & 0.46 & 0.56  \\
Black \& van Dishoeck model S2		&
	0.08 & 0.03 & 0.21 & 0.08  \\ 
\hline
\end{tabular}
\end{center}
\end{table}

\clearpage

\begin{table}
\begin{center}
\caption[]{Br$\gamma$ Fluxes in NGC 7714}
\label{BRlines.tab}

\myvspace

\begin{tabular}{ccc} \hline \hline

Aperture & Flux & Source$^a$ \\
~ & (10$^{-18}$\,W\,m$^{-2}$) & \\
(1) & (2) & (3) \\ \hline

$1.22\arcsec\times 2.44\arcsec$	& 21.0 $\pm$ 0.6	& $a$ \\
$1.22\arcsec\times 6.10\arcsec$	& 27.5 $\pm$ 0.7	& $a$ \\
$3.5\arcsec\times 7\arcsec$	& 26.4 $\pm$ 3.8	& $b$ \\
$3.1\arcsec\times 9.3\arcsec$	& 29.5 $\pm$ 2.2	& $c$ \\
$6\arcsec\times 6\arcsec$		& 30 $\pm$ 3		& $d$ \\
$7\arcsec$ diam.		& 47 $\pm$ 5		& $e$ \\
$10\arcsec\times 20\arcsec$	& 49.9 $\pm$ 2.0	& $f$ \\
$10.3\arcsec\times 20.7\arcsec$	& 51 $\pm$ 12		& $g$ \\
\hline
\end{tabular}
\end{center}

\myvspace

$a$- This paper;
$b$- Taniguchi et al. (1988);
$c$- Puxley \& Brand (1994);
$d$- Moorwood \& Oliva (1988);
$e$- Ho et al. (1990);
$f$- Calzetti (1997);
$g$- Kawara et al. (1989).

\end{table}

\end{document}